\documentclass[manuscript]{aastex}
\usepackage{natbib}
\usepackage{graphicx}

\shorttitle{Thickening of the thin disk in the Third Quadrant}
\shortauthors{R.\ A.\ V\'{a}zquez et al.}

\begin{document}


\title{The thickening of the thin disk in the third Galactic quadrant\thanks{Based on observations carried out at Cerro Tololo Inter-American Observatory, under programs CHILE-0008B-017 and CHILE-0010A-006}}


\author{Giovanni Carraro\altaffilmark{1}}
\affil{European Southern Observatory, Alonso de Cordova 3107,
Casilla 19001, Santiago 19, Chile}
\email{gcarraro@eso.org}

\author{Rub\'{e}n A.\ V\'azquez}
\affil{Facultad de Ciencias Astron\'omicas y Geofísicas (UNLP), Instituto de Astrof\'isica
de La Plata (CONICET, UNLP), Paseo del Bosque s/n, La Plata, Argentina}
\email{rvazquez@fcaglp.unlp.edu.ar}

\author{Edgardo Costa}
\affil{Departamento de Astronom\'ia, Universidad de Chile,
Casilla 36-D, Santiago, Chile}
\email{costa@das.uchile.cl}

\author{Javier A.\ Ahumada}
\affil{Observatorio Astron\'omico, Universidad Nacional de C\'{o}rdoba, Laprida~854,
5000~C\'ordoba, Argentina}
\email{javier@oac.uncor.edu}

\and
\author{Edgar E.\ Giorgi}
\affil{Facultad de Ciencias Astron\'omicas y Geof\'{\i}sicas (UNLP), Instituto de Astrof\'{\i}sica
de La Plata (CONICET, UNLP), Paseo del Bosque s/n, La Plata, Argentina}
\email{egiorgi@fcaglp.unlp.edu.ar}

\altaffiltext{1}{On leave from Dipartimento di Astronomia,
Universit\`a di Padova, Italy.}

\begin{abstract}

In the third Galactic quadrant ($180\degr \leq l \leq 270\degr$)
of the Milky Way, the Galactic thin disk exhibits a significant
warp ---shown both by gas and young stars--- bending down a few
kpc below the formal Galactic plane ($b=0\degr$). This warp shows
its maximum at $l\sim240\degr$, in the direction of the Canis
Major constellation. In a series of papers we have traced the
detailed structure of this region using open star clusters,
putting particular emphasis on the spiral structure of the outer
disk. We noticed a conspicuous accumulation of young star clusters
within 2-3 kpc from the Sun and close to $b=0^{o}$, that we
interpreted as the continuation of the Local (Orion) arm towards
the outer disk. While most clusters (and young stars in their
background) follow closely the warp of the disk, our decade-old
survey of the spiral structure of this region led us to identify
three clusters, Haffner~18(1 and 2) and Haffner~19, which remain very close
to $b=0^{\degr}$ and lie at distances (4.5, $\sim$8.0,  and 6.4 kpc) where most of the material
is already significantly warped.

Here we report on a search for clusters that share the same
properties as Haffner~18 and 19, and investigate the possible
reasons for such an unexpected occurrence. We present \emph{UBVRI}
photometry of 5~young clusters, namely NGC~2345, NGC~2374,
Trumpler~9, Haffner~20, and Haffner~21, which also lie close to
the formal Galactic plane. With the exception of Haffner~20, in
the background of these clusters we detected young stars that
appear close to $b=0\degr$, and are located at distances up to
$\sim 8$ kpc from the Sun, thus deviating significantly from the
warp. These populations define a structure that distributes over
almost the entire third Galactic quadrant. We discuss this
structure in the context of a possible thin disk flaring, in full
similarity with the Galactic thick disk.

\end{abstract}


\section{Introduction}
\label{sec:intro}

The relevance of the study of the open star clusters of the Milky
Way is unique for many areas of modern astronomy such as stellar
astrophysics, stellar and Galactic evolution, and Galactic
structure,  and is widely recognised as such.

With this motivation, during the last decade we have been securing
homogeneous \emph{UBVRI$_{kc}$} observations (same telescope,
set-up, and strategy of data analysis) of young open clusters in
the third Galactic quadrant (3GQ: $180^{\degr} \leq l \leq
270^{\degr}$) of the Milky Way, in order to study its complicated
structure. In this fashion, we have contributed significantly to
the present understanding of the spiral structure in this Galactic
region (Carraro et al. 2005, Moitinho et al. 2006, V\'azquez et
al. 2008, Carraro et al. 2010).

Recent summaries of the status of this investigation and of the
questions that still remain unanswered are given in Carraro (2014)
and Carraro et al. (2014). Briefly, and using Fig.~1 as reference,
our extensive campaign showed that the Local (Orion) arm (LOA) is
an inter-arm feature, a bridge most probably connecting the
Carina-Sagittarius arm in the first Galactic quadrant with the
outer (Norma-Cygnus) arm in the 3GQ. The LOA also apparently
breaks the Perseus arm, which is not prominent at all in young
stars or gas in this region of the Galaxy. The extension of the
LOA into the 3GQ occurs towards $l \sim 240^{\degr}$, that is in
the direction of the Canis Major over-density ($b \sim -8\degr$,
Martin et al. 2004; Momany et al 2006; Moitinho et al.
2006). This picture confirms independent findings based on HI
studies in the same region (Levine et al. 2006). Because our
sample extends up to 20~kpc from the Galactic center, we were also
able to trace the progressive bending of the thin disk caused by
the warp, and did not find any indication of a disk cutoff or
truncation at 14~kpc from the Galactic center, as previously
postulated by Galactic models (Robin et al. 1992).\\

\noindent
Two of the main open questions that arise from Fig.~1 are: (i) at what
distance from the Sun the disk starts to bend down, and  (ii) whether
flaring (the thickening of the disk in the vertical direction),
clearly visible in the old disk population (Momany et al. 2006, L\'opez-Corredoira \&  Molg\'o 2014),
has some impact on the structure of the Galactic thin disk as
well.

Our previous studies indicate that the warp becomes important
beyond 12~kpc from the Galactic center ($\sim4$~kpc from the Sun).
From the works of V\'{a}zquez et al. (2008, Fig.~6), and
V\'{a}zquez et al. (2010, Fig.~15), it is evident that the disk
keeps close to the formal ($b = 0^{\degr}$) plane up to about
12~kpc, and that it is very thin ($\pm100$~pc ). Beyond this
distance, the thin disk starts bending, and at the same time its
vertical extent gets more and more significant. Although the bulk
of young star clusters follows the warp closely, there is a group
of young star clusters, like Haffner~18(1/2)  and Haffner~19 ( at 4.5 , $\sim$8, and 6.4 kpc, see
Fig.~1) that clearly and un-expectedly deviate from the main disk
locus. Although these clusters are relatively nearby, 
we found young stellar populations in their
background, at distances up to 9~kpc from the Sun, which also do
not follow the warp. These latter might be indicating a flaring of
the thin disk, or, more conservatively, they could be tracing
local irregularities (e.g. disk corrugations) of the same kind as
the Big~Dent (Alfaro et al. 1991, Moitinho et al 2006).

\noindent
In this paper we report on an extensive search for other
presumably young clusters in the same Galactic sector, and close
to the formal Galactic plane, like Haffner~18 and 19.. The aim of
this study is twofold: we want (1) to improve the basic parameter
of a set of poorly studied, but potentially interesting clusters,
and (2) to use them to consolidate our knowledge of the thin disk
structure in the 3GQ. Specifically,  we present results for five
open clusters (see Table~1): NGC~2345 (C0706$-$130), NGC~2374
(C0721$-$131), Haffner~20 (C0754$-$302), Haffner~21 (C0759$-$270),
and Trumpler~9 (C0753$-$258), based on new \emph{UBVRI$_{kc}$}
data.\\

\noindent
Since all these five clusters have been very poorly studied, a
significant part of this paper is dedicated to the description of
the data and the derivation of their properties. Therefore, the
outline of the paper is as follows.  In Section~2 we present a
brief literature overview of the 5 clusters; in Section~3 we
present the observations, the data reduction procedure and the
photometry; in Section~4 we make a comparison of our data-set with
previous investigations; in Section~5 we describe the star count
procedure and derivation of the cluster's sizes; and in Section~6
we present the analysis of the clusters' photometric diagrams. In
Section~7 we discuss the properties and spatial distribution of
early type field stars in the clusters back- and fore-ground.
Finally, in Section~8 we summarise the main conclusions of our
work.

\section{Literature overview}

A search of the literature shows that the data for the five clusters are truly scanty, and that no modern
(CCD) studies are available.

\noindent {\bf NGC 2345}:\\
 NGC 2345 is an open cluster projected towards  the
Monoceros constellation. Moffat (1974) carried out photoelectric
photometry of 62~stars down to $V=14$. He detected
variable reddening ($E_{(B-V)}=0.5$ to $\sim 1.10$), and placed
the cluster at a distance of 1.75~kpc. By means of star counts,
Moffat found that the cluster diameter is 10.5~arcmin, but
indicated that, using only faint stars, it could be as large as
12.3~arcmin. Interestingly for the purpose of the present study,
Moffat also isolated a small number of blue stars which he
classified as OB by means of objective prism observations, and
that could be at distances as large as 10~kpc. Lastly, from the
inspection of the photometric diagrams he inferred for the cluster
an age of $\sim 6\times10^{6}$~yr.

\noindent {\bf NGC~2374}:\\
 Like NGC~2345, this cluster is projected towards the Monoceros
constellation. Photographic and photoelectric photometry for
39~stars was secured by Babu (1985), who estimated a uniform color
excess $E(B-V) = 0.175$, an age of $75 \times 10^{6}$~yr,
and a distance of 1.2~kpc. We note that the age derived by Babu
(1985) is significantly younger than previous estimates, such as
those from Fenkart et al. (1972) and Lyng\aa\ (1980), who report
ages of $3.5\times10^{8}$ and $2\times10^{9}$~yr, respectively.

\noindent {\bf Trumpler~9}:\\
Trumpler 9 is a Puppis association object.
Pi{\c{s}}mi{\c{s}} (1970) found, from photoelectric and
photographic data, that the earliest spectral type is A0, that its
reddening in $(B-V)$ is at most 0.2~mag, and that its
spectroscopic distance is approximately 900~pc. Vogt \& Moffat
(1972) used photoelectric \emph{UBV}-H$\beta$ photometry and
derived a larger reddening $E(B-V) = 0.29$, an apparent distance
modulus of 11.98, and a distance of 1.62~kpc. Their estimated
bluest spectral type is B1, very different from
Pi{\c{s}}mi{\c{s}}'s results.

\noindent {\bf Haffner~21 and 20}:\\
These two clusters also belong to Puppis
association. Based on photographic photometry, Fitzgerald \&
Moffat (1974) established B9 as the earliest spectral type present
in Haffner 20 and Haffner 21. They also determined a distance and
reddening of 2.4~kpc and 0.55~mag for Haffner~20, and of
3.3~kpc {and 0.20~mag for Haffner~21, respectively.

\section{Observations, data reduction and photometry}

In this Section we briefly introduce the observation strategy, the collected data,
and the technique employed to reduce them. Since this work is part of a long series of papers, 
full details about the observations, data reduction procedure, and
the photometry can be found in previous papers from our group
(see, e.g., Carraro et al. 2010a,b).

\subsection{Observations}

Observations were carried out with the Y4KCAM camera attached to
the Cerro Tololo Inter-American Observatory (CTIO, Chile) 1-m
telescope, operated by the SMARTS consortium\footnote{\tt
http://http://www.astro.yale.edu/smarts}, in February 2008 and
December 2010. This camera is equipped with an
STA~$4064\times4064$
CCD\footnote{\texttt{http://www.astronomy.ohio-state.edu/Y4KCam/detector.html}}
with 15-$\mu$m pixels, yielding a scale of
0.289$^{\prime\prime}$/pixel and a
field-of-view (FOV) of $20^{\prime} \times 20^{\prime}$ at the Cassegrain focus of the telescope.\\

In Table~2 we present the log of our \emph{UBVRI} observations.
All observations were carried out in photometric, good-seeing
conditions. Our \emph{UBVRI} instrumental photometric system was
defined by the use of a standard broad-band Kitt Peak
\emph{UBVRI$_{kc}$} set of
filters.\footnote{\texttt{http://www.astronomy.ohio-state.edu/Y4KCam/filters.html}}
To determine the transformation from our instrumental system to
the standard Johnson-Kron-Cousins system, and to correct for
extinction, each night we observed Landolt's area SA~98 (Landolt
1992) multiple times (3 or 4), and with different air-masses ranging from
$\sim1.1$ to $\sim2.2$. Field SA~98 includes over 40 well-observed
standard stars, with a good magnitude and color coverage: $9.5\leq
V\leq15.8$, $-0.2\leq(B-V)\leq2.2$, $-0.3\leq(U-B)\leq2.1$.

\subsection{Reductions}

Basic reduction of the CCD frames was done using the
Yale/SMARTS y4k reduction script based on the IRAF\footnote{IRAF
is distributed by the National Optical Astronomy Observatory,
which is operated by the Association of Universities for Research
in Astronomy, Inc., under cooperative agreement with the National
Science Foundation.} package \textsc{ccdred}, which includes bias and
sky-flats correction. 
Photometry
was then performed using IRAF's \textsc{daophot} and \textsc{photcal}
packages. Instrumental magnitudes were extracted following the
point spread function (PSF) method (Stetson 1987) using a
quadratic, spatially variable master PSF (PENNY function). We typically use 
a minimum of 50 PSF stars spread over the detector.
Finally, the PSF photometry was aperture corrected using aperture
corrections determined making aperture photometry of bright,
isolated stars in the field.

\subsection{The photometry}

To transform our instrumental magnitudes to the standard system we used equations of the form:
\begin{equation}
M = m + c_0 + c_1 \times X + c_2 \times (B-V),
\end{equation}

where $M$ and $m$ are the standard and instrumental magnitudes,
$X$ is the airmass, and $c_0$, $c_1$, $c_2$ are the zero point,
first-order extinction, and color term coefficients, respectively.
We note that the use of second-order extinction terms did not
improve the fit. In Table~\ref{tab:coeficientes} we list the
coefficients of the transformation equations, and their
corresponding errors, for each night.

In this way we produced calibrated \textit{UBVR$I_{kc}$}
photometric catalogs for NGC~2345, NGC~2374, Haffner~20,
Haffner~21, and Trumpler~9, containing 6326, 3266, 3740, 6725, and
6119 entries, respectively.

These optical catalogs were cross-correlated with 2MASS (Skrutskie et al. 2006) to
convert  pixel (i.e., detector coordinated)  into equatorial  RA and DEC for equinox J2000.0,
thus providing 2MASS-based astrometry.

\section{Comparison with previous studies}

As mentioned previously, no modern CCD studies exist for the five
clusters we are presenting in this study, so we want to note that
we are making a comparison of our CCD data just with photographic
and photoelectric data. In general, photoelectric data are
precise, but suffer from two major problems: they are not very
deep, and in crowded fields they cannot resolve blends because of
the relatively large fixed apertures adopted. On the other hand,
photographic data are a in general somewhat deeper, but are
less precise, and quickly loose efficiency near the sensitivity
limit because of poor quantum efficiency.

Figs.~2 and~3 show the results of the comparison; below we comment
on a cluster-by-cluster basis. We note that all mean differences
presented were computed in the sense our measurements minus others
and errors on the mean are standard deviations.

\noindent {\bf NGC~2345}:\\
 Photometric differences with photoelectric measures for
62~stars in common with Moffat (1974) are shown in the upper
panels of Fig.~2. Mean differences are: $\Delta V =
-0.086\pm0.07$, $\Delta (B-V) = -0.015\pm0.03$, and $\Delta (U-B)
= +0.04\pm0.08$.
\noindent
While colors are in fine agreement, our $V$ magnitudes are
somewhat brighter, but still comparable within the errors.
Moffat explains that  his observations were carried out using
a small telescope.  Therefore we believe this may lead to probable
contamination in the star light due to the presence of faint
neighbours in the diaphragm area and/or difficulties in removing
the sky contribution onto the star light. These are the typical limitations associated
with photo-electric photometry. In this respect, CCD
imaging combined with PSF photometry allow  more precise removal of the contamination
sources and also a better isolation of the star light.

\noindent {\bf NGC~2374}:\\
 Photometric differences for 36~stars in common with Babu
(1985) are shown in the lower panels of Fig.~2. If we take all the
36~stars into account (photoelectric and photographic data taken
together), the mean differences are: $\Delta V = -0.16\pm0.16$,
$\Delta (B-V) = +0.108\pm0.23$, and $\Delta (U-B) = +0.11\pm0.22$.
\noindent
 Excluding the 10~stars with photographic photometry (all fainter than $V \sim 14.0$), the mean
differences become:
$\Delta V = -0.15\pm0.09$,
$\Delta (B-V) = -0.01\pm0.05$, and
$\Delta (U-B) = +0.03\pm011$,
\noindent
In this last case the scatter around the means
significantly decreases and the color agreement is fine. However,
our $V$ magnitudes still appear slightly brighter with large
scatter. We could not find any detailed information either about the quality of
the photoelectric observing runs by Babu (1985) [made at the
Kevalur 102-cm telescope in India] or related to the errors involved in the
measures, but we draw the attention to Babu who suggests
photographic mean errors above 0.2 for $B$ and $V$ and even larger
in $U$. This said, one can  understand the zero point offset
between Babu photoelectric magnitudes and
ours.

\noindent {\bf Trumpler~9}:\\
The photometric comparison is shown in the
upper panels of Fig.~3. Empty triangles depict the comparison with
the photographic data from Pi{\c{s}}mi{\c{s}} (1970), while the
comparison to her photoelectric data is depicted by filled
circles. Filled squares show the comparison against the
photoelectric photometry by Vogt \& Moffat (1972).
The comparison with photographic photometry from
Pi{\c{s}}mi{\c{s}} (1970) yielded: $\Delta V = +0.17\pm 0.08$,
$\Delta (B-V) = -0.09\pm0.13$, and $\Delta (U-B) = -0.06\pm0.18$
for 13 stars in common.  Both the offset in $V$ o
and the scatter around the mean are strong. Something worse happens when comparing
with photoelectric measures for 10 stars also from
Pi{\c{s}}mi{\c{s}} (1970). In this case we found the following
differences: $\Delta V = -0.50 \pm 0.64$, $\Delta (B-V ) = 0.01
\pm 0.08$, and $\Delta (U-B) = 0.01 \pm 0.12$. The mean  $\Delta
V$ is absolutely unrealistic. Such a curious mean difference
leads us to conclude that some of her photoelectric $V$ measures
are simply wrong. There is also, in our opinion, a
propagation effect producing the large $V$ off-set with
photographic and photoelectric photometry since they come from bad
data calibration by Pi{\c{s}}mi{\c{s}} (1970). In brief, she used
this wrong photoelectric sequence to put her photographic data
into the standard system. This scenario is supported by the comparison with
 Vogt \& Moffat (1972) photometry.
We found 20 stars in common with their photoelectric measures. As
seen in Fig. 3, there is a significant discrepancy in $V$ for 
six of them. 
Once these these six problematic stars have been removed, the
mean differences with Vogt \& Moffat (1972) are: $\Delta V = +0.05
\pm 0.06$, $\Delta (B-V ) = 0.00 \pm 0.04$, and $\Delta (U-B) =
0.03 \pm 0.12$. The $V$ offset is relatively small and well within
the error. The six deviating stars can be either mis-identified, or variable, objects.\\

\noindent {\bf Haffner~20 and 21}:\\
The comparison against the photographic photometry of
Fitzgerald and Moffat (1974) is shown in Fig.~3; in the middle
panels for Haffner~21 (47~stars), and in the bottom
panels for Haffner~20 (31~stars). Mean differences are:
$\Delta V = -0.09\pm0.08$, $\Delta (B-V) = +0.05\pm0.09$, and
$\Delta (U-B) = +0.03\pm0.15$ for Haffner~21. For Haffner~20:
$\Delta V = +0.06\pm0.09$, $\Delta (B-V) = +0.06\pm0.11$, and
$\Delta (U-B) = -0.13\pm0.14$.
In both clusters most of the scatter comes from the faint
tail of the photometry, as one can readily see inspecting Fig.~3.
However, and particularly in the case of Haffner 20, there is a
distinctive group of largely deviating stars, some of them for
more than 0.8 mag in $V$. 
These stars are distributed over the whole
 magnitude range,  which makes it  impossible to
separate, in the photographic plates, images of close stars and
obtain their true photometric values. 
It must also be added that the calibration of photographic magnitudes by Fitzgerald and
Moffat (1974) was done using (secondary) photoelectric standards
from some nearby open cluster sequences. Fitzgerald and
Moffat (1974) mention that the external errors of their
calibration is of the order of 0.09 in $V$, $B$ and $U$ filters.
As for the internal standard errors of the mean they claim they
are all above 0.05 in $V$, $B-V$ and $U-B$. This large
uncertainty in the old photographic measures can easily explain
the differences with our photometry.

\section{Star counts and clusters's size}

The linear size is a fundamental quantity for  a
comprehensive characterisation of a star cluster. It is necessary
to estimate its evolutionary status through the analysis of its
density profile and other related parameters (see Aarseth 1996; de La
Fuente Marcos 1997; Kroupa et al. 2001) as the slope of the
mass function. It is also routinely used as a criterion  for
separating cluster-dominated from field-dominated regions in
photometric diagrams (Baume et al. 2004).  

The usual way  to estimate the radius of a star clusters 
is to assume a centrally peaked spherical stellar density distribution and determine the
distance at which the density converges to the mean field background .
This determination is done either
by visually setting the limit in a plot or by fitting King (1962) or other
profiles to the run of radial density. However, both approaches
assume a spherically symmetric distribution and, in the case of
the King profiles, it also assumes that the system is dynamically relaxed. 
This is not at all the case for Galactic ope clusters.\\

\noindent
Since we are mostly interested in determining the cluster radii to reduce field
contamination in photometric diagrams and derive the best possible
cluster parameters, the following approach has therefore been adopted:
contour maps were computed in each field by choosing adequate
kernel sizes and grid steps. Once a notorious stellar density is
identified, a circle enclosing its full range is drawn by eye and
the cluster limit is taken as that defined by the radius of this
circle. We adopt the cluster center to be the geometric center of
the cluster as opposed to the position of the density peak. We
want to emphasise that unlike what usually happens with globular
clusters, the center of an open cluster can rarely be
unambiguously identified. Furthermore, assigning a unique center
value to an open cluster with complicated non-circular morphology
or poorly populated may often carry unclear implications of
what precisely that center represents, in terms of the cluster's
structure.

Fig.~4 shows contour maps of the five clusters under analysis.
The insert in each figure, the  colour-coding  represents the 
number of stars per arcmin$^2$. With the exception of NGC~2374,
we detect in all cases a clear over-density around the nominal
cluster coordinates. As for NGC~2374, if the cluster exists, it
might simply be larger than the area we covered with our
photometry since no clear over-density is appreciable in Fig. 4.
NGC~2345 and Haffner~21 emerge clearly as star clusters, but their
regions suffer from variable extinction, exemplified by the
conspicuous dust lanes, in both cases in the north-west area of
the corresponding images. The most evident peak is that of
Haffner~20, while Trumpler~9 shows several peaks, mostly in the
northern region. The central peak in Trumpler 9, however,
is evident enough to consider it as the star cluster since it
coincides with the cluster location in earlier studies and we use
it to set its limits. We recall the readers that our method is
intended to just yield an approximation to the cluster size
because of the loose and asymmetric nature of open star clusters,
and the often wrong  impression of an over density induce by
variable visual absorption. 
This is particularly important for Trumpler
9, Haffner 21 and NGC 2345 where the variable extinction in the
stellar field  introduces important
fluctuations in the star density background which, in turn, alters
the cluster extension.\\

\noindent
We then used star counts to derive the radial density profile, and
measure the cluster radial extent, improving over previous, mostly
visual, estimates.  The results are shown in Fig.~5. In all cases,
a distinct over-density is detected, with radii of $\sim$ 4.0,
2.5, 4.0, 2.0, and 1.0 arcmin for NGC~2345, NGC~2374, Trumpler~9,
Haffner~21, and Haffner~20 (from top to bottom of Fig.~5)
respectively. In the case of NGC~2374, the radius pf 1.4 arc min refers to the
central over-density as seen in Fig.~4 and commented on in Sect~.6.

\section{Analysis of the clusters' photometric diagrams}

To derive the basic parameters of the star clusters under
study, we have performed a detailed inspection of the two color
(TCD) and color-magnitude (CMD) diagrams after having verified its
reliability and defined the angular cluster sizes according to the
results found in Section 5. The next step in the analysis consists
in the identification of the cluster sequences in all the
photometric diagrams, determine the color excesses by shifting the
Schmidt-Kaler (1982) to fit the sequences in the TCD and estimate
the distance by superposition of the ZAMS onto the blue envelope
of the corresponding star sequence in the CMD. In each case,
both the fit and its error have been estimated by eye on a 
trial and error basis. 
Once mean reddening and distance have been determined we
 fit isochrones (Marigo et al. 2008) computed with solar
metallicity. Again, this fitting has been performed by eye but
instead of error in the fitting we suggest a range of values. As
for the value of the absorption law (the R-value) in each cluster
we have adopted $A_V= R\times E_{(B-V)}$ with $R=3.1$. Our
adoption of this $R$-value is supported in the long series of
articles see for instance, Moitinho (2001), Carraro et al. (2007)
or V\'azquez et al. (2010). we have performed in this quadrant of
the Galaxy where we have demonstrated, using the $B-V,V-I$
color-color diagrams, that the absorption law is normal with very few
exceptions, as Haffer 18/19 discussed in V\'azquez et al. (2010).
Finally, we reiterate once again our methodology is a classical and well-known
procedure (see Straizys 1991 for an exhaustive review).
Error analysis has been conducted following Carraro et al. (2007) and Carraro (2011) propagation formulae, which
take into account the effects of spectral mis-classification on distance and reddening.\\
In the following we only comment on the results, on a
cluster-by-cluster basis.

\noindent {\bf NGC~2345}:\\
 The TCDs and CMDs shown in the upper panels of
Fig.~6 were constructed using all stars within 3.75~arcmin from
the cluster center. The TCD shows a clear and slightly
blue sequence partially contaminated by field interlopers as seen
in the corresponding CMDs. Indeed, a superficial view of the CMD can
lead to the wrong conclusion that variable reddening is present
amongst the stars in NGC 2345 in line with earlier findings from
Moffat. However,  a careful inspection of the position of
these stars in the TCD and CMDs simultaneously is enough to conclude that
most of the stars causing the spread are, indeed, foreground
stars. The fit of  the intrinsic color relation (Zero Age Main
Sequence, ZAMS) from Schmidt-Kaler (1983) to the mean cluster
sequence in the TCD (upper left panel) provides a reddening
$E_{(B-V)} = 0.59\pm0.04$.  Absorption is obviously highly
variable across the cluster. This is evident in Fig. 4 (upper left
panel) where the number of stars decreases strongly to the
north-west. However, the effect of variable reddening is not so strong in the cluster area,  in
agreement with Fig. 6 and the low dispersion around the mean color
excess. In the CMD (middle upper panel in Fig. 6) we
show the same ZAMS fitted vertically to the blue envelope
of the cluster sequence, which yields an apparent distance
modulus $(V-M_V)  = 14.2\pm0.1$.  When removing the effects
of visual absorption onto the apparent distance modulus we derive
a distance of $3.0\pm0.5$~kpc from the Sun for NGC 2345. A
similar fitting to the $V$ vs. $(V-I)$ CMD (right upper panel)
confirms previous findings. In both CMDs, an estimate of the age
is obtained by superposing the cluster sequence with
models from the Padova group (Marigo et al. 2008). The
super-imposed isochrones  bracketing the cluster sequence are for
log(age) = 7.8--7.9 (63--70~Myr).\\

\noindent
From our analysis, NGC~2345 turns out to be significantly more
distant than the 1.75~kpc given by Moffat (1974).
Such a large difference in the cluster distance comes from
the fact that Moffat's data only reach $V \sim 14$ mag, a
magnitude level where confusion with field interlopers is
particularly high (see Figs. 3 in Moffat paper). At the
level of the magnitude cut-off of Moffat data he had no
chance to see the run of  the lower MS sequence as we instead do in Fig. 6.
Therefore,  our ZAMS fitting is much more robust because  it encompasses
an almost 6 mag range,  down to $V \simeq 18$.\\

\noindent
An interesting, final,  comment is necessary for the sake of clarity and its relevance 
for the main aim of this work.
A close inspection of the lower panels of Fig.~6,
where supposedly only field stars are present, shows that several
stars fall between the reference lines for stars of spectral types
O4 and~B5. Because these stars have unique reddening solution in
the TCD, we applied the Q-method (Strai\v{z}ys 1991) to extract
their intrinsic colors, color excesses, and distances, as we have
done successfully many times in the past for several other fields (see e.g.,
Carraro et al. 2007; V\'{a}zquez et al. 2010; Carraro et al 2010).
We will return to these stars again in Section~7.

\noindent {\bf NGC~2374}:\\
In the previous Section we noted that this cluster is
not obvious at all from the contour plot in Fig. 4 upper
right panel. However, when looking at the upper panels in Fig.~7,
it is quite evident the presence of a slightly evolved
sequence in the CMD. The earliest spectral type is
$\sim$~B9. This sequence extends for more than five magnitudes in
the CMDs and is affected by a uniform reddening $E_{(B-V)} =
0.07\pm0.03$. Fitting a ZAMS yields an apparent distance
modulus of $(V-M_V) = 10.90\pm0.10$, which, after
correcting by visual absorption, places this cluster at a distance
of $1.33\pm0.25$~kpc, in agreement with the estimate of Babu
(1985). In Fig~7 (upper panels) we have over-plotted isochrones
from the Padova group (Marigo et al. 2008) on the $V$ vs. $(B-V)$
and $V$ vs. $(V-I)$ CMDs. They indicate that  the cluster is
220--280~Myr (log(age)~$=$~8.35--8.45). This age range
indicates that the cluster is older than assumed by Babu (1985),
younger than the estimates from Lynga (1980),
$2\times 10^9$ yr, and comparable to the age obtained by Fenkart
et al (1972), $3.5\times10^8$ yr.

As we have already mentioned,  NGC 2374 is a sparse open cluster
and probably it covers a larger area than the one we have
surveyed. We adopt as field stars those ones that do not belong to
the cluster sequence we have defined in the TCD and CMD upper
panels. Inspection of the TCD and CMDs for the stellar field in
the direction of NGC~2374 (lower panels in Fig.~7) shows the
presence of a population of stars (indicated by red symbols) which
defines a Blue Plume (BP). In other words, the BP stars do not share the
average locus occupied by field stars but they are blue-wards of
the field sequence intersecting it  at $V \sim 17-18$ mag.

We remind the reader that previous studies from our group have
shown that these stars, the BP stars, do not constitute a
physical sequence of bound stars, but a superposition of stars
along the line of sight (Carraro et al. 2005). We will return
to these stars in the following section.

\noindent {\bf Haffner 20}:\\
This cluster stands out notoriously in the contour plot of Fig. 4.
Visual inspection of the TCD and CMD (upper panels of Fig. 8) for
stars within the adopted cluster radius ( r$ < $ 2 arcmin)
confirms that Haffner 20 is a physical group. The Main Sequence
(MS) of this cluster is quite clear in the TCD, and is affected by
a uniform reddening of E(B -V ) = 0.65 $\pm$0.03. Still, like in the
case of NGC 2345, contamination by field interlopers is
evident in the cluster area, although top a lesser degree.
 Fitzgerald and Moffat (1974) give a maximum $E_{(B-V)}$
for this cluster of about 0.55. Regrettably, they do not show the
corresponding TCD of this object (nor the one of Haffner 21) but
we have redrawn the TCD (not shown for saving space) with their
data finding a strong data dispersion that makes impossible to
assign a trustable reddening to Haffner 20 (this is fully in line with
our photometric comparison in Section 4). The apparent distance
modulus ($V-M_V$) that we infer from ZAMS fitting of the CMD is
$15.70\pm0.30$, which, in combination with the reddening estimated
above, places the cluster at a distance of $5.5\pm 1.0$~kpc from the
Sun. This new value is twice the obtained by Fitzgerald and
Moffat (1974) and is the result of the use of a deep photometric
dat-set that allows to follow the lower cluster MS and perform
a more solid ZAMS fitting. The earliest spectral type, as estimated
from the TCD, is about B5-B7, which suggests an age around
100~Myr. This is supported by the two isochrones from Marigo et
al. (2008), for log(age) between 7.95 and 8.00, that we
over-plotted on the CMDs. This range lowers the
cluster age with respect the 200 My suggested by Fitzgerald and Moffat
(1974). The field toward Haffner 20 does not reveal any presence
of a young population in the background. There might be various explanations for such an occurrence.
One possibility is that the background population is extremely reddened, and we do not detect it because it
is confused with the cluster and field MSs. Another possibility is that there is indeed no background
populations. Several times in previous works (see e.g. Carraro et al. 2005) we show cases of presence and 
absence of background stellar populations. 

\noindent {\bf Haffner 21}:\\
A similar analysis for Haffner~20 leads to an estimate
of the reddening of $E(B-V) = 0.21\pm0.03$ (see Fig.~9, upper left
panel). The MS for the probable members of Haffner~21 is well
defined in both the TCD and the CMD, and indicates that Haffner~21
may be slightly older than Haffner~20, because its earliest
spectral type would be around B8--B9. From the CMDs (Fig~9, upper
right panels) we infer an apparent distance modulus $(V-M_V) =
13.35\pm0.35$, which puts the cluster at a distance of
$3.5\pm0.5$~kpc from the Sun. In this case, our reddening
and distance are in agreement with previous studies from Fitzgeral
and Moffat (1974). The superposition of isochrones from Marigo et
al. (2008) indicates that the age range for log(age) is between
8.05 and 8.1 or 1.1 to $1.25\times 10^8$ yr, significantly younger than the
$2\times 10^8$ yr found by Fitzgeral and Moffat (1974).\\

\noindent
In the control field (lower panels of Fig.~9, in red symbols) we
detect a young population in the background of the cluster. It is
not possible to position accurately this population, because it
seems to be composed of stars located at very different distances,
from close to the Sun to all the way up to about 10~kpc from it.
We further analyze this population in Section~7.

\noindent {\bf Trumpler 9}:\\
The photometric diagrams for the stars within the
cluster radius (see Fig.~10, upper panels) indicate that we are
dealing with an extremely young object. The MS in all the diagrams is
very tight, and the reddening solution provides a color excess
$E(B-V) = 0.20\pm0.02$, with no spread.  By fitting the CMDs with
a ZAMS we derive an apparent distance modulus $(V-M_V)=
12.90\pm0.30$. This yields an heliocentric distance of
$2.9\pm0.5$~kpc, \a bit larger than the values from Vogt \&
Moffat (1972) and Pi{\c{s}}mi{\c{s}} (1970). This distance is
consistent with Trumpler~9 being part of the Puppis~OB1
association, which is also corroborated by our estimate of the
cluster age ($\sim 10$~Myr) from the isochrone fitting.
\noindent
A glance at the lower panels in Fig.~10 allows us to conclude that
there is also an evident young stellar population in the
background of Trumpler~9, as in the case of Haffner~21.
Moreover, from the TCD one can conclude that young stars (red
symbols) show a wide range of reddening, which conservatively
implies that they are located at very different distances, and
hence they are tracing an elongated structure in this particular
Galactic direction (see below).

\subsection{Trumpler 9, Haffner 20 and Haffner 21 and their relation to
the Puppis association}

These three clusters are located in a region of the Galaxy
historically identified as Puppis constellation, and that  contains two 
prominent OB
associations: Puppis OB1 and Puppis OB2. The approximate borders of the area
are between $l^{\circ} \sim 242$ and  $l^{\circ}\sim 246$
and from $b^{\circ}\sim +2$ to $b^{\circ}\sim-1$. 
Although the nature and properties of 
these two OB associations is out of the scope of this
work,  still we would like to  comment on the relationship of Trumpler~9, and Haffner 20 and 21
with them.\\

\noindent
We first of all emphasise that no clear understanding of this region is available.
Humphrey (1978) found various components of what is designated as Puppis OB1 association at
2.5 kpc from the Sun,  while  Havlen (1976) first reported a second, more distant, OB
association, Puppis OB2 at 4.3 kpc. The existence of these two
separate associations have been questioned by Kaltcheva and Hilditch (2000),
who could not find evidences of them, but, instead,  suggested the presence  in the same
Galaxy direction of  two other star groups (at 1 kpc the first, and 3.2 the second), 
but significantly lower, at $b^{\circ}\sim-4$. These would
surround the star cluster NGC 2439, located  at 3.5-4.5 kpc from the Sun.\\

\noindent
Since Trumpler 9 is located at 2.9 and
it is $10^6$ yrs old,  we argue it is a probable distant member of
Puppis OB1.
On the other hand,  Haffner 20, at 5.5 kpc and about $10^8$ yrs old, would be
placed in front of Puppis OB2 and be  an old member
of this association. Something similar happens with Haffner 21. It distance of 
3.5 and its  comparable age to Haffner~20 would make it  a distant member
of Puppis OB2. \\

Lastly, let us emphasise that the assessment  of these three clusters membership to
these associations (even if they really exist) requires an extensive
spectroscopic and radial velocity study. In line with
Kaltcheva and Hilditch (2000) study,  the existence of two young
stellar groups at $b^{\circ}\sim-4$ and with a distance range of 1
to 3.2 kpc in addition to the more distant NGC 2439 open cluster is clearly telling us
that star formation processes is still vigorous in this region
of the Galaxy.

\section{Early-type field stars in the fore- and back-ground of the clusters}

Groups of evenly distributed early-type stars are frequently found
in the background of Galactic open clusters (see, for instance,
Carraro et al. 2005) in the third Galactic quadrant. This occurs because Galactic open clusters
are mostly located at low Galactic latitudes in this Milky Way regions, and the line of sight
to them often intersects parts of distant spiral arms. These stars
manifest themselves in classical photometric diagrams in two ways.
First, we can spot them in the \emph{UBV} TCD in the position
corresponding to stars for which a unique reddening solution is
derived (Strai\v{z}ys 1991). The reddening solution is achieved
moving the stars along the reddening lines back to the position of
blue-types on the intrinsic (zero reddening) line (see Perren et
al. 2012, for a vivid example and a detailed description of the
method). Second, in the CMD they tend to define tight sequences,
emerging from the left (the blue) side  of the more prominent MS of
the thin (or thick) disc stars. These features are well known as
BPs (Carraro et al. 2005; Moitinho et al. 2006), and their exact
location in the CMD depends on the line of sight.
In this study, field blue stars have been detected in NGC~2345
(Fig~6, lower panels) and Trumpler~9 (Fig.~10, lower panels),
while BPs are evident in NGC~2374 (Fig.~7, lower panels) and
Haffner~21 (Fig.~9, lower panels). In what follows we discuss the
properties of these groups of stars individually.

\subsection {NGC~2345}
NGC~2345 and NGC~2374 are only 4 degrees apart on sky (see Table~1)
FNGC~2345 is a young open cluster located at 
intersection of the Perseus arm with the LOA. The TCD of stars
outside the limits of NGC~2345 (lower panels of Fig.~6) shows a
large number of stars above the reference line for stars of
spectral type~B5, with different reddening values roughly along the color
range $-0.1 \leq (B-V) \leq1.0$. As we mentioned in
Section~6, field blue stars in NGC~2345 were de-reddened using the
Q-method to derive their intrinsic colors $(B-V)_{0}$ and individual
color excesses $E(B-V)$. We remind the reader that once these two
quantities are obtained for a star, a spectral type and an
absolute magnitude can be assigned using the relationships given
in Schmidt-Kaler (1982). This procedure is a variation of the
spectroscopic parallax method applied to get individual star
distances, that has been proven to be quite solid (see, e.g.,
Carraro et al. 2007; Carraro et al. 2010).\\

\noindent
In Fig.~11 we show the trend of reddening versus distance for the
four fields in which blue stars have been detected. To minimise
the effects of an erroneous spectral type assignment due to
photometric errors, we have only taken into account stars with
spectral types earlier than~B4. In the case of NGC~2345, Fig.~11
shows the existence of large $E(B-V)$ values, from 0.5 to~1.2,
from the Sun to about 3.5~kpc, distance at which the reddening
levels off at a mean value of $E(B-V) = 1.2 \pm0.2$.

Most  blue stars in the field of NGC~2345 are found south-ward the
cluster. At a distance of 3.0~kpc, the reddening 
is larger than the reddening of the cluster itself
($E(B-V) = 0.59$), as if the cluster were seen through a dust
window. We note that, as shown by the contour plot presented in
Fig.~4, that absorption significantly increases towards the north-west
of the cluster. 

Lastly, from Fig.~12 one  can notice that these early-type stars
are found at any distance up to 8~kpc, and that a few of them lie
at distances larger than 10~kpc. All of them are located below the
formal $b=0\degr$ plane,  roughly following the trend defined by the
majority of  young open clusters in the first 3~kpc from the Sun.

\subsection{NGC~2374}
In the CMD of NGC~2374 a clear BP emerges at $V\sim17$. This BP
corresponds to the group of stars with $0.0 \leq (B-V) \leq 0.5$
in the TCD (lower panels of Fig.~7), where we have depicted in red
about 50 BP~candidate stars. The bulk of these BP stars has a
reddening value of $0.40\pm0.10$ (Fig.~7, left panel), and an
apparent distance modulus of $15.2\pm0.2$,  or a distance of
$6.1^{+0.7}_{-0.5}$~kpc (Fig.~7, middle and right panels).

This BP structure clearly deserves closer attention, so we applied
the same method for determining individual distances and color
excesses to the 50 BP candidate stars that we used for the blue
stars in the field of NGC~2345. Inspection of Fig.~11 shows in
this case that:

\begin{description}
\item - BP stars begin to appear at $d\sim1.5$~kpc and extend for more than 8~kpc;
\item - color excess increases steadily from 0.25 to 0.5;
\item - the earliest photometric spectral type found among the BP stars in NGC~2374 corresponds to B5-type stars,
while latest types are around A0.
\end{description}

At odds with NGC~2374 itself , early-type stars in its background 
are confined above the formal Galactic disk plane.

\subsection{Haffner~21}
The TCD of field stars in the direction of Haffner~21 (lower left panel of Fig.~9) reveals the
presence of a strip of stars immediately below the O4~reddening line, from $(B-V)= 0.4$ to 1.2.
They are depicted by black symbols.\\
Haffner~21 exhibits  a notorious BP (see Fig.~9, lower panel)
extending for more than five magnitudes. We marked some of the
stars in the BP with red symbols and fitted them with the
Schmidt-Kaler ZAMS. This procedure yielded an apparent distance
modulus of $15.2\pm0.4$ that, for a reddening value of
$0.30\pm0.12$, corresponds to a distance of
$7.1^{+1.4}_{-1.2}$~kpc. We also applied the Q-method to obtain
the intrinsic properties of these stars which we assume are
representative of the BP. Fig.~11 shows that these stars are
distributed starting at 1.5~kpc from the Sun, and that the
reddening steadily increases up to a value of about 0.25.

\subsection{Trumpler~9}
Trumpler~9 is at 1.3 degrees above the galactic plane and at 1.8
degrees away from Haffner~21; it is also above the galactic plane.
These two clusters are located close to the northern edge of the
Vela Gum SN~remnant.
As in the case of NGC~2345, in the field of Trumpler~9 there is an
evident group of blue stars outside the cluster's boundary
(depicted by red symbols in Fig.~10, lower panels). In the TCD
they define a strip between the reddening lines for 04- and
A0-type stars, thus admitting a unique reddening solution. This
situation is reflected in the CMDs (Fig.~10, mid and right lower
panels), where BP stars are colour-coded in red as well.
From our analysis, the earliest photometric spectral type found
among these stars is about B0.5; Fig.~11 shows that their
reddening $E(B-V)$ has a mean value of $0.20\pm0.10$, which stays
constant up to 4~kpc. 
Beyond 4~kpc, a few stars show $E(B-V)$  values near 0.9 and helio-distance up to 8 kpc.
An outlier - having large reddening- is at 2 kpc from the Sun. 
Both these group are located north-east of Trumpler~9, the most 
absorbed region covered by our images.

\section{Discussion and Conclusions}
In this work we have presented new \emph{UBVRI} CCD data for five
poorly known Galactic star clusters:  NGC~2345, NGC~2374,
Haffner~20, Haffner~21, and Trumpler~9. Our data supersede
previous  photographic and photoelectric studies, both in depth
and  quality, which allowed us to constrain the clusters's basic
parameters more solidly. We report in Table~4 the results of this
new analysis, in comparison with previous studies.\\

\noindent 
We stress, however, that the main goal of this work is to study
the spatial distribution of early-type field stars that often
happen to lie in the background of open clusters in the 3GQ of the
Milky Way (Moitinho et al. 2006; V\'azquez et al 2008). To summarise our conclusions along this vein, 
we refer the reader to Fig.~12 ---which is an updated version of Fig.~1---,
where we have added the new findings that have resulted from the
analysis of the cluster sample presented in this paper.

\noindent
 Before interpreting our data, we would like to further stress that our sample is mostly made
of {\it bona fide}  OB stars. In fact, one way wonder that our sample can be significantly contaminated by 
old hot sub-dwarf stars, or that the OB stars we detect out of the plane are mainly OB run-away stars.
We would like to comment on these two possibilities.\\

\noindent
Hot sub-dwarf stars are often found in the Milky Way. They are associated with old stellar populations, and they are in most cases binaries.  The statistics of these stars are poorly known. Early estimates  (Downes 1986) count one hot sub-dwarf every ten million stars, and one B sub-dwarf every million stars, when considering one cubic parsec volume.
Recent studies (Breedt \& Gansicke 2011, and references therein) confirm these figures. These stars are very faint, and, when detected nearby to the Sun, can be confused with distant OB stars, in absence of any spectroscopic information. 
The typical magnitude of these stars is M$_{V} \approx +5.0$. This means that with our limiting magnitude of $V \approx 17.5$ for the BPs, we can detect them at a distance modulus up to  $(m-M) \approx 12.0$, assuming no reddening. This turns into a distance from the Sun of about 3.0 kpc, thus defining a volume of $\sim$ 340,000 cubic parsec,
since our detectors covers 20 $\times$ 20  arcmin$^2$ on the sky.  This implies that we expect 0.2 O sub-dwarf per cubic parsec,
and 0.7 B sub-dwarf per cubic parsec (according to Downes 1986) in each of our pointings . An additional point is that OB sub-dwarf stars do not
define the typical tilted BP sequences we have found, which, on the other hand, resemble more the MS of nearby young open clusters.\\

\noindent
OB run-away stars are as well often found in the Milky Way disk. They are typically associated with massive young open clusters(e.g. Westerlund~2, Carraro et al. 2013; Roman-Lopes et al. 2011), from which they escaped (Fujii \& Portegies Zwart 2011). First of all we stress that the escaping direction is not always vertical with respect to the disk plane, but it is clearly random. In Westerlund~2 we believe there are two run-away OB stars out of about 50 OB stars in the cluster. No OB run-away stars have been reported in young open clusters in the anti-center direction, most probably because clusters are not massive enough to produce this phenomenon (Fujii \& Portegies Zwart 2011),
which indeed is mostly seen in the nearby of massive star cluster located within the solar ring.
Even assuming that the proportion of run-away stars in the outer disk is the same as in the inner, say about 
20$\%$, it is difficult to reproduce the high number of off-plane OB stars we are finding.\\

\noindent
In Fig.~12, which covers the whole 3GQ, from 180$\degr$ to
270$\degr$ in Galactic longitude, we depict the positions of the
young open clusters studied by our group, and of over-densites of
young stars that we have discovered in their background or
foreground. In this figure the Sun is located at $(0,0,0)$. Black
stars are young open clusters from our previous studies (Moitinho
et al. 2006, V\'{a}zquez et al. 2008, and V\'{a}zquez et al.
2010), the two black stars surrounded by a blue circle are
Haffner~18 and Haffner~19, and the five red stars are the clusters
analyzed in this work.  Crosses mark the distribution of
early-type stars in the directions of NGC~2345 (green), NGC~2374
(brown), Trumpler~9 (red), and Haffner~21 (black), respectively.
Open circles are BPs detected in previous studies. The solid
logarithmic curves are the approximate extrapolated locations of
the Perseus and Outer arms, according to  Valle\'e (2005) model.

\noindent
The addition of the five clusters studied here has added a wealth
of new information on the structure of the Galactic thin disk in
the 3GQ, thanks to the fact that four of  them (the exception
being Haffner~20) lie in fields rich in early-type stars located
at very different distances from the Sun.

NGC~2345 and the associated blue stars follow the well-established
structure of the disk in the 3GQ: the disk keeps close the formal
$b=0\degr$ Galactic plane up to $\sim$ 4~kpc from the Sun, and
then starts to bend down following the Galactic warp. However,
field stars in the direction of NGC~2374, Trumpler~9 and
Haffner~21 are found all the way up to the location of the outer
arm, and, being these clusters at positive Galactic latitudes,
they define distributions of young stars that, at odds with  any
previous expectations, lie above or close to  the formal
$b=0\degr$ Galactic plane. In other words, as illustrated in the
lower panel in Fig.~12, they do not follow the Galactic warp.
Beyond $\sim 4$~kpc from the Sun, the Galactic thin disk, as
defined by young stars and open clusters, thickens significantly,
and this thickening keeps increasing because of the sudden
appearance of the Galactic warp.

\noindent 
We interpret this situation as an indication that the Galactic
thin disk is not only warped, but also flared, like the thick disk
(Momany et al. 2006; L\'opez-Corredoira \& Molg\'o 2014).  In other words, the Galactic thin disk
scale height seems to increase moving outwards from the Sun. This
interpretation nicely explains the unusual, un-expected location
of Haffner~18 and Haffner~19 (see Fig.~12): they are very young (NGC~2374, Trumpler~9, and Haffner~21)
and distant, but lie far from the main -warped- thin disk
location. The addition of  background blue stars- associate with three more clusters (T-- sharing the same position as
Haffner~18 and 19 --- lends further support to this scenario.

\noindent 
One may wonder whether other independent observational evidences
of this thickening of the disk are available, in addition to the
few stellar fields discussed here. We note that in their recent
all-sky survey of HII~regions in the Wide-Field Infrared Survey
Explorer (WISE) database, Anderson et al. (2014) detected quite a
large number of HII~regions in the Galactic anti-center. Their
Fig.~4 shows very clearly both the warp and the flare of the thin
disk in the second and third Galactic quadrants ($90\degr \leq l
\leq 270\degr$), nicely confirming our findings.\\

\noindent 
To conclude, we have shown in  our series of papers,
including the present one, that early type stars and very young
open clusters not only follow the warp in the 3GQ,  but they also
contribute to the flaring of the thin disk beyond $\sim$4 kpc from
the Sun. This might imply that in the outer disk spiral arms are
thicker than commonly believed. In fact a large fraction of the
stars we have studied are younger than 100 Myr, hence  it is
conceivable to think that they did not displace much  from their
birthplaces.Ê This implies  that these young stars formed out of
vertically dispersed material, which is not concentrated  into the
plane, but can be found even at large distances from the mean
plane.\\

\noindent
This interpretation is supported by the evidence (Carraro et al
2005, V\'azquez et al. 2008) that, besides early type stars, also
CO clouds are found at any vertical distances in the 3GQ quadrant,
and  their spatial distribution  closely  young stars and star
clusters. In a very recent study, Suad et al. (2014) have
performed a search of HI super-shells in the second and third
quadrant of the Galaxy showing that HI super-shells can be found
up to 3 kpc above the galactic plane (50\% of them are at less
than 500 pc above the galactic plane). Early type stars,  young
clusters, CO clouds and HI super-shells show all the same smoothed
distribution that thickens the outer Galactic disk.

\acknowledgements
GC thanks Yazan Momany for many fruitful discussions.
RAV and JAA acknowledge ESO for granting a visitor-ship at ESO
premises in Santiago, where part of this work was done. EC
acknowledges support by the Fondo Nacio\-nal de
Investiga\-ci\'{o}n Cient\'{\i}fica y Tecnol\'{o}gica (Project
No.~1110100 Fondecyt), and the Chilean Centro de Exce\-lencia en
Astro\-f\'{\i}sica y Tecno\-log\'{\i}as Afines (PFB06).
RAV and EEG acknowledge financial support from the PIP
1359 from CONICET. This research used the WEBDA database,
maintained at the Institute for Astronomy of the University of
Vienna. Lastly, we greatly appreciate the input of the referee.

%%%%%%%%%%%%%%%%%%%%%%%  FIGURES %%%%%%%%%%%%%%%%%%%%%%%%%%%%%%%%%%%

\clearpage
\begin{figure}
\epsscale{0.8}
\plotone{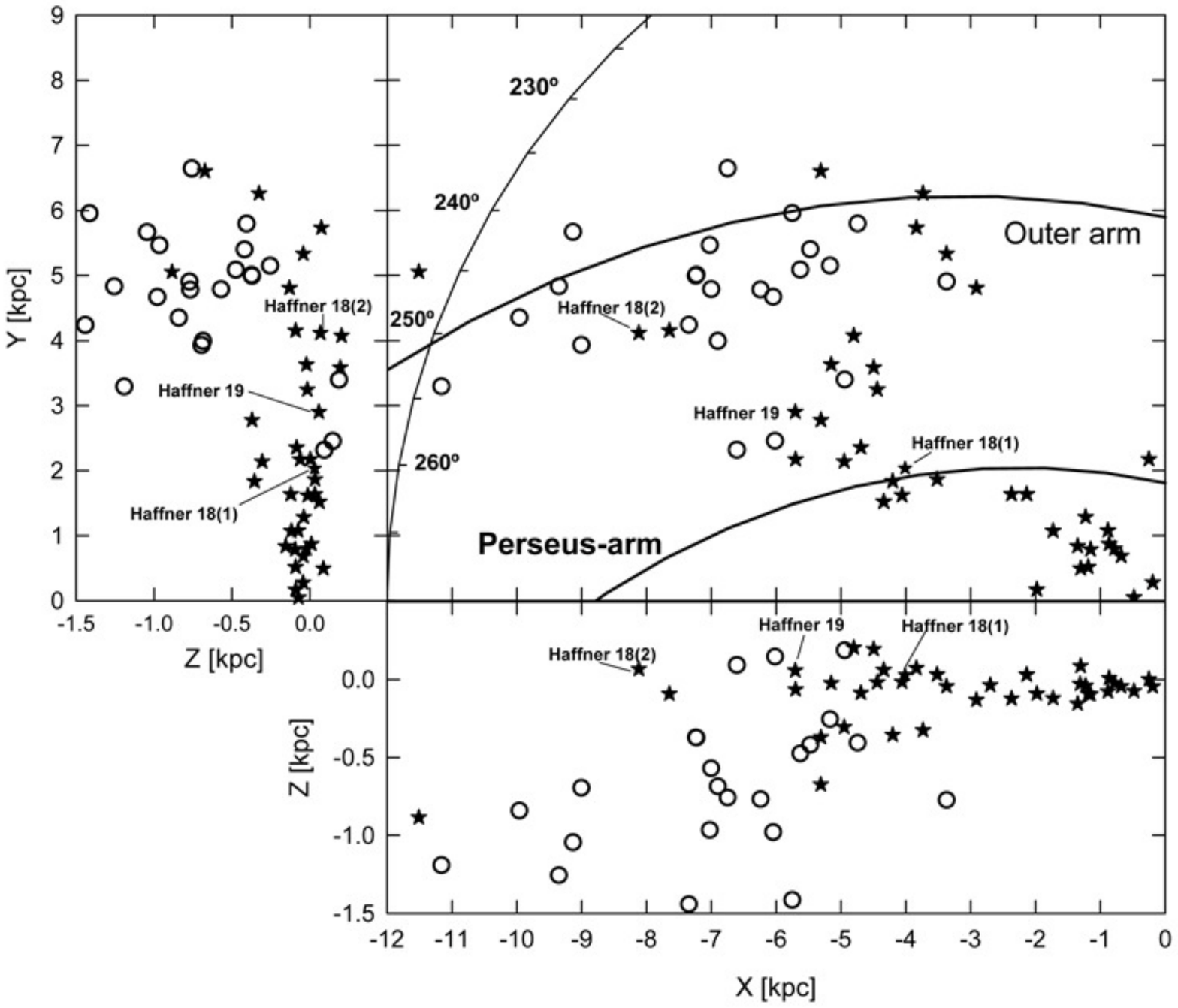}
\caption{
The thin-disk structure in the third Galactic quadrant as traced by young open clusters and background populations.
The large panel shows the plane of the disk, while the two smaller panels show the \emph{X--Z} and \emph{Y--Z}
projections. Notice the
particular location of the star clusters Haffner~18(1/2)  and~19, above the formal ($b=0\degr$) plane of the Galaxy. Symbols are as follows:
stars are young open clusters, and empty circle background populations of early type stars.}
\end{figure}

\clearpage
\begin{figure}
\epsscale{1.}
\plotone{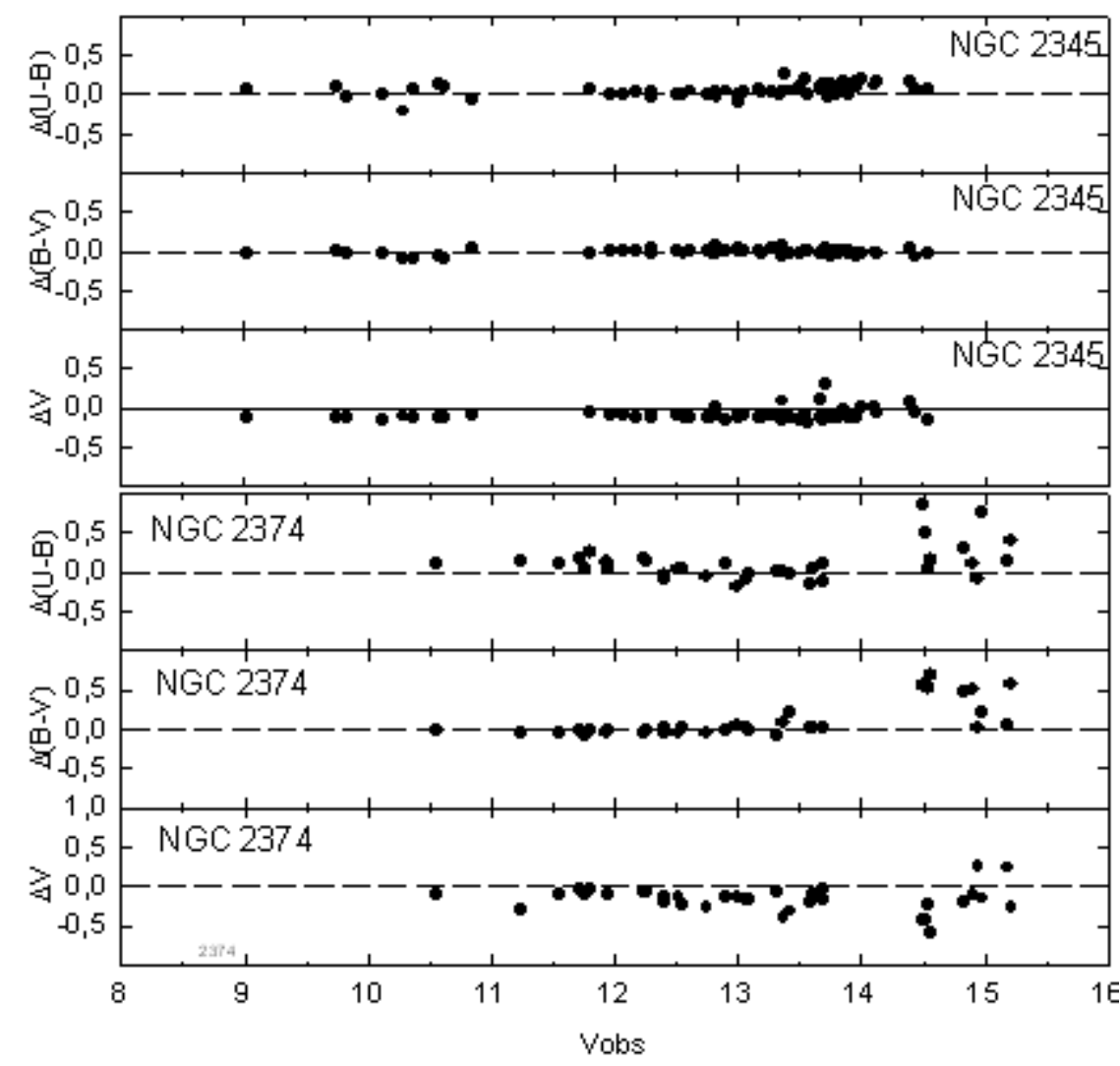}
\caption{Comparison of our CCD photometry with literature sources for NGC~2345 and NGC~2374. Differences are in the
sense: our measurements minus published data. See text for details.}
\end{figure}

\clearpage
\begin{figure}
\epsscale{0.8} \plotone{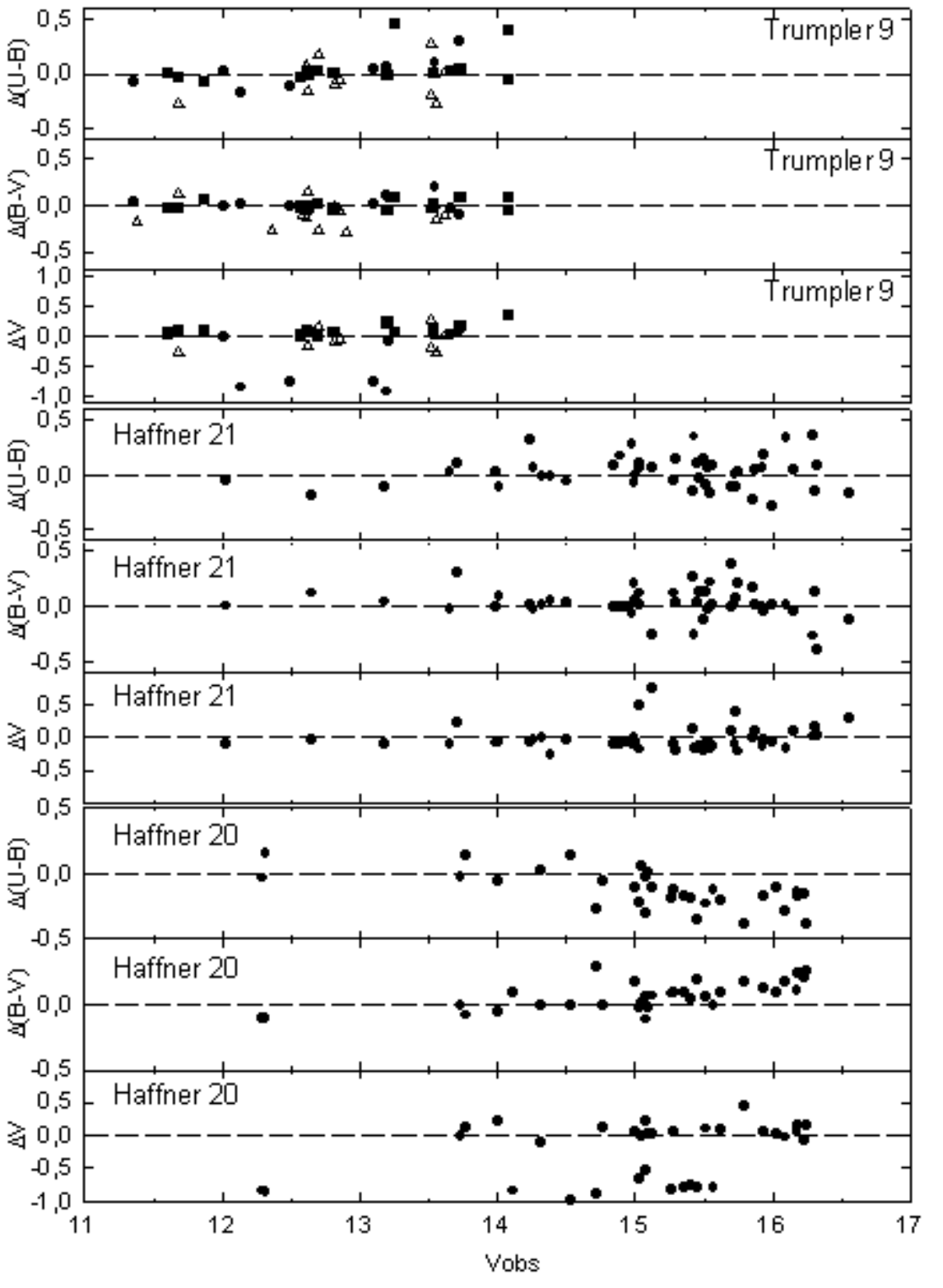} \caption{Comparison of our CCD
photometry  for Trumpler~9, Haffner~20, and Haffner~21.
Differences are in the sense: our measurements minus published
data. See text for symbols in Trumpler 9. }
\end{figure}

\clearpage
\begin{figure}
\epsscale{0.8}
\plotone{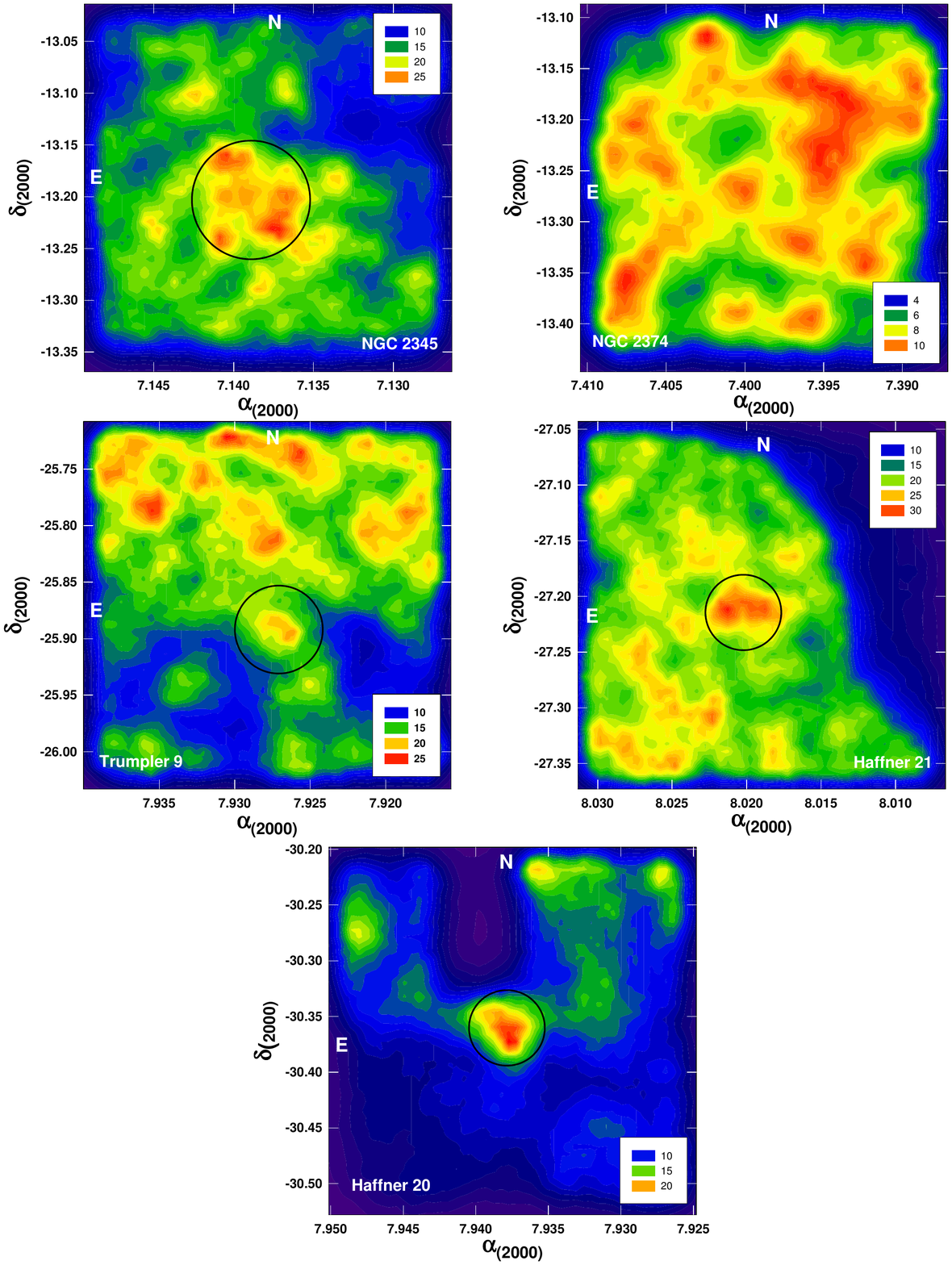}
\caption{Iso-density contour maps for the five clusters studied.}
\end{figure}

\clearpage
\begin{figure}
\epsscale{0.8}
\plotone{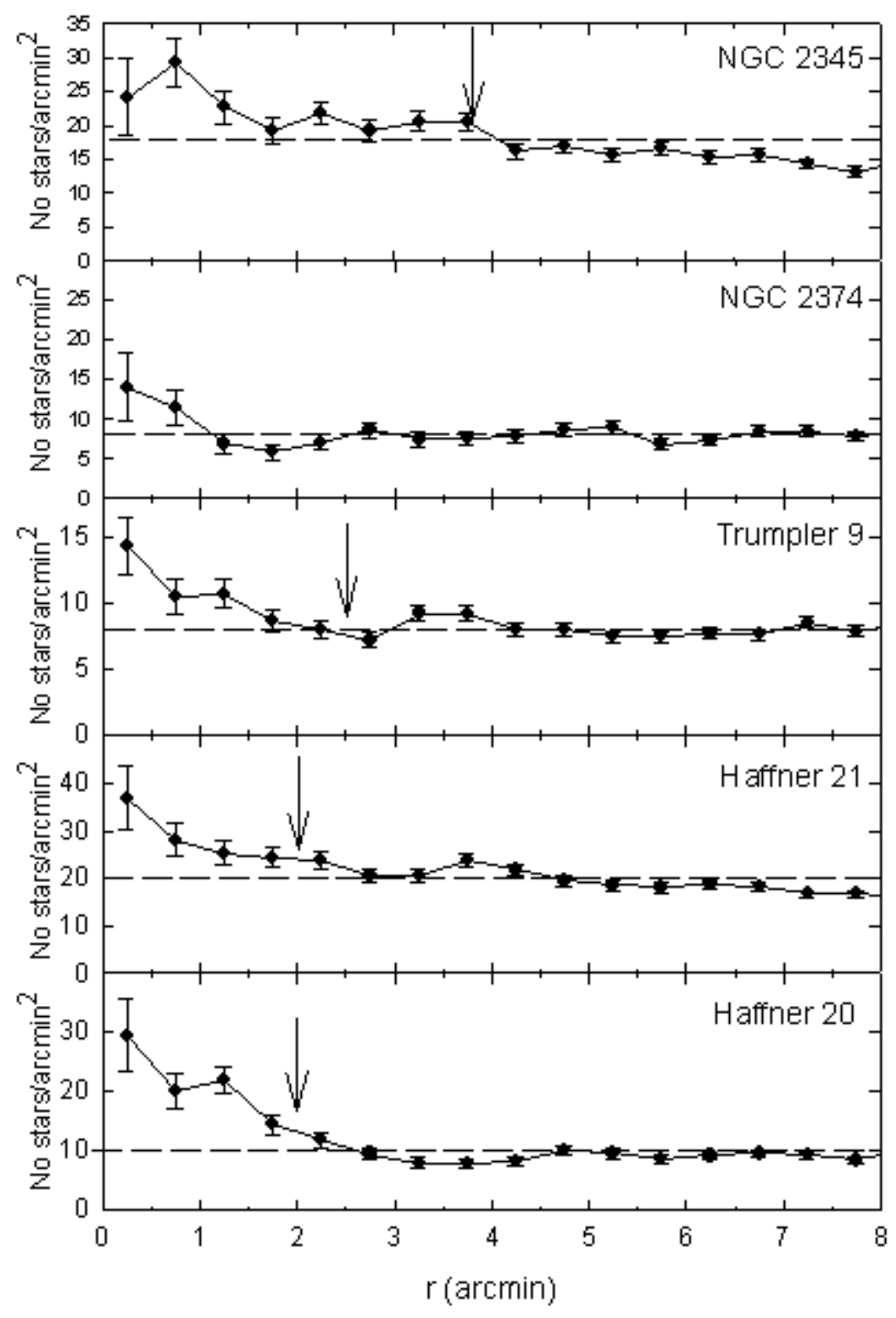}
\caption{Star counts as a function of radius. See text for details.}
\end{figure}

\clearpage
\begin{figure}
\epsscale{1.} \plotone{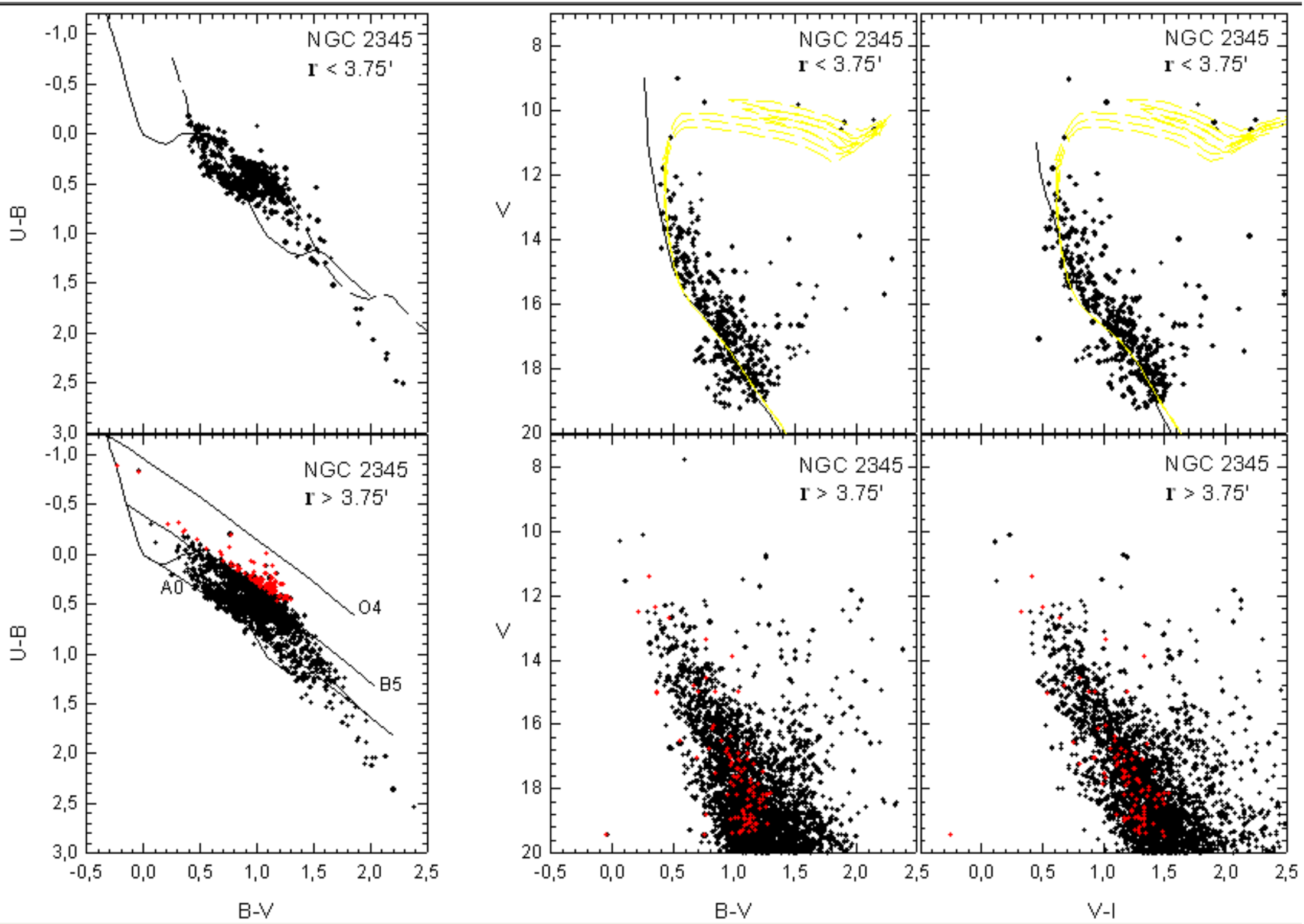} \caption{Color-color and
colour-magnitude diagrams for NGC~2345 (upper panels), and for the
surrounding field (lower panels). Dashed lines are isochrones from
Marigo et al. (2008) shifted by color excess and distance modulus.
The solid line in the two color diagram is a reddening-free ZAMS,
while the dashed line is the same ZAMS shifted by the color
excess. Two reference spectral types are indicated in the lower
left panel, together with the reddening vector. See text for
details.}
\end{figure}

\clearpage
\begin{figure}
\epsscale{1.} \plotone{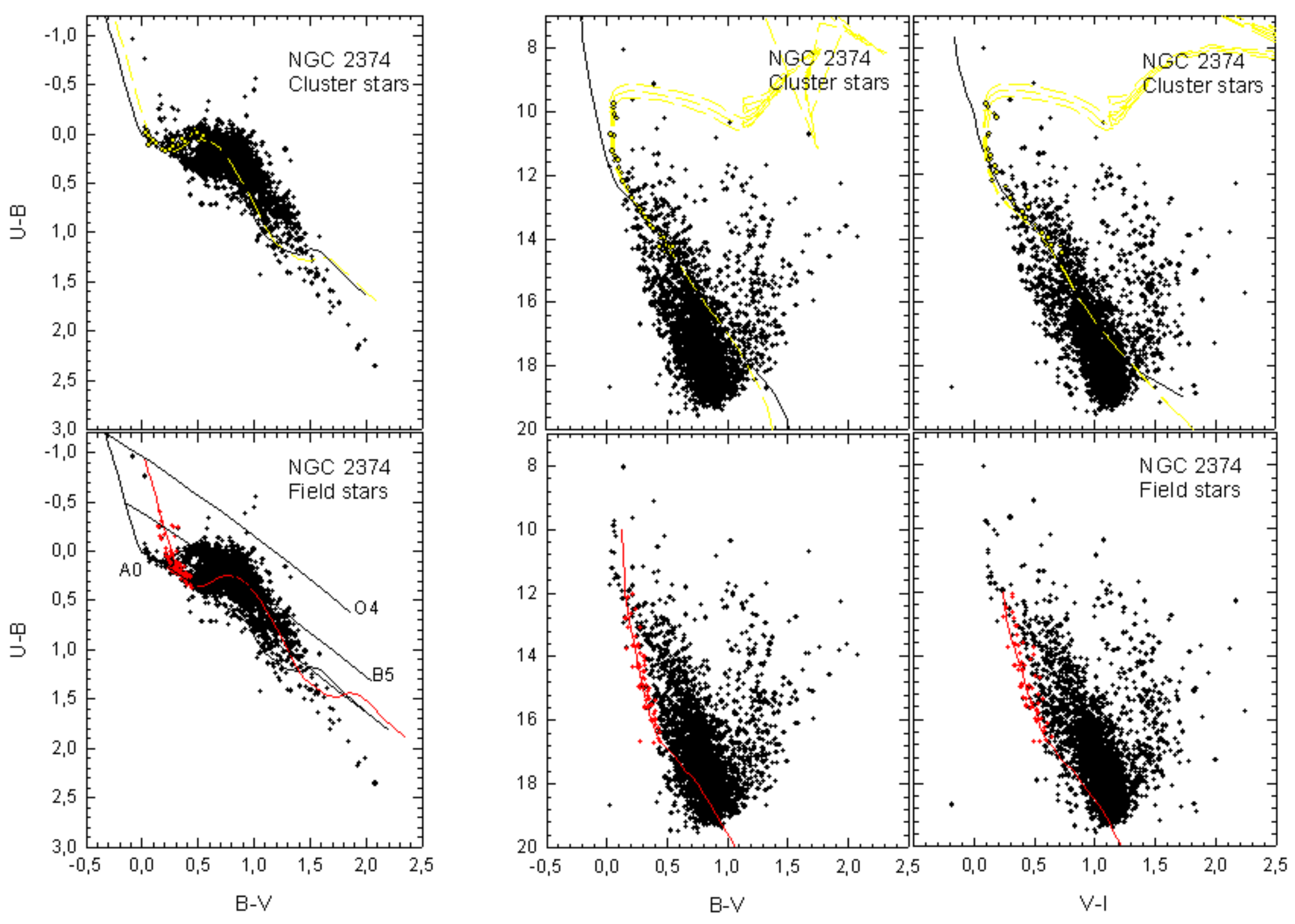} \caption{Color-color and
colour-magnitude diagrams for NGC~2374 (upper panels), and for the
surrounding field (lower panels). Yellow dashed lines are
isochrone from Marigo et al. (2008) shifted by color excess and
distance modulus. The solid line in the two color diagram is a
reddening-free ZAMS, while the dashed yellow line is the same ZAMS
shifted by color excess to fit the cluster's probable members
(filled yellow symbols). In the lower panels, red colours are used
to identify a background young population, and the ZAMS that
better fits its sequence. See text for details.}
\end{figure}

\clearpage
\begin{figure}
\epsscale{1.} \plotone{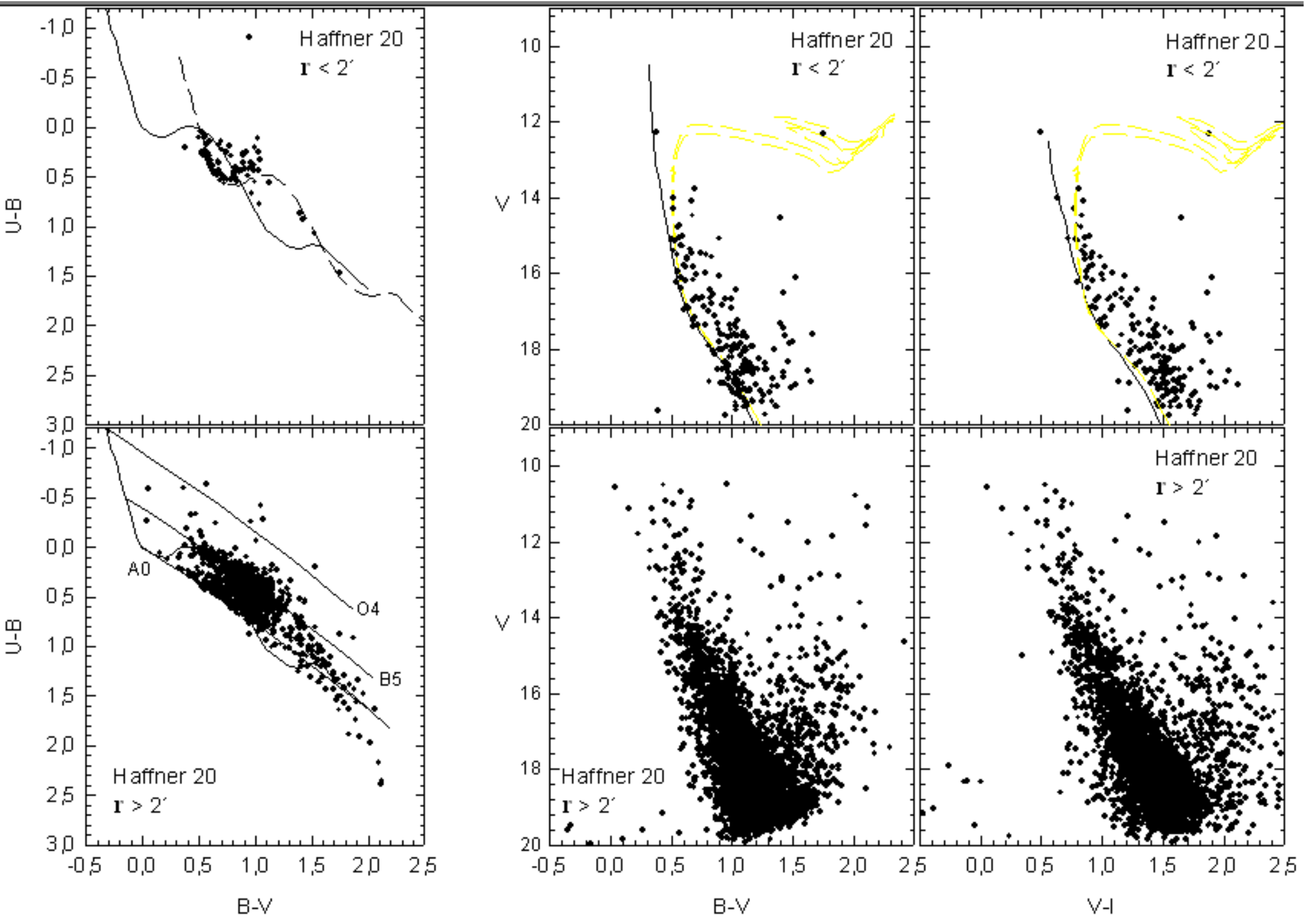} \caption{Color-color and
color-magnitude diagrams highlighting  Haffner~20 members (upper panels), and 
the surrounding OB field stars (lower panels). Dashed lines are isochrone
from Marigo et al. (2008) shifted by color excess and distance
modulus. The solid line in the two color diagram is a
reddening-free ZAMS, while the dashed  line is the same ZAMS
shifted by the color excess to fit the cluster's probable members.
Two reference spectral types are indicated in the lower left
panel, together with the reddening vector. See text for details.}
\end{figure}

\clearpage
\begin{figure}
\epsscale{1.} \plotone{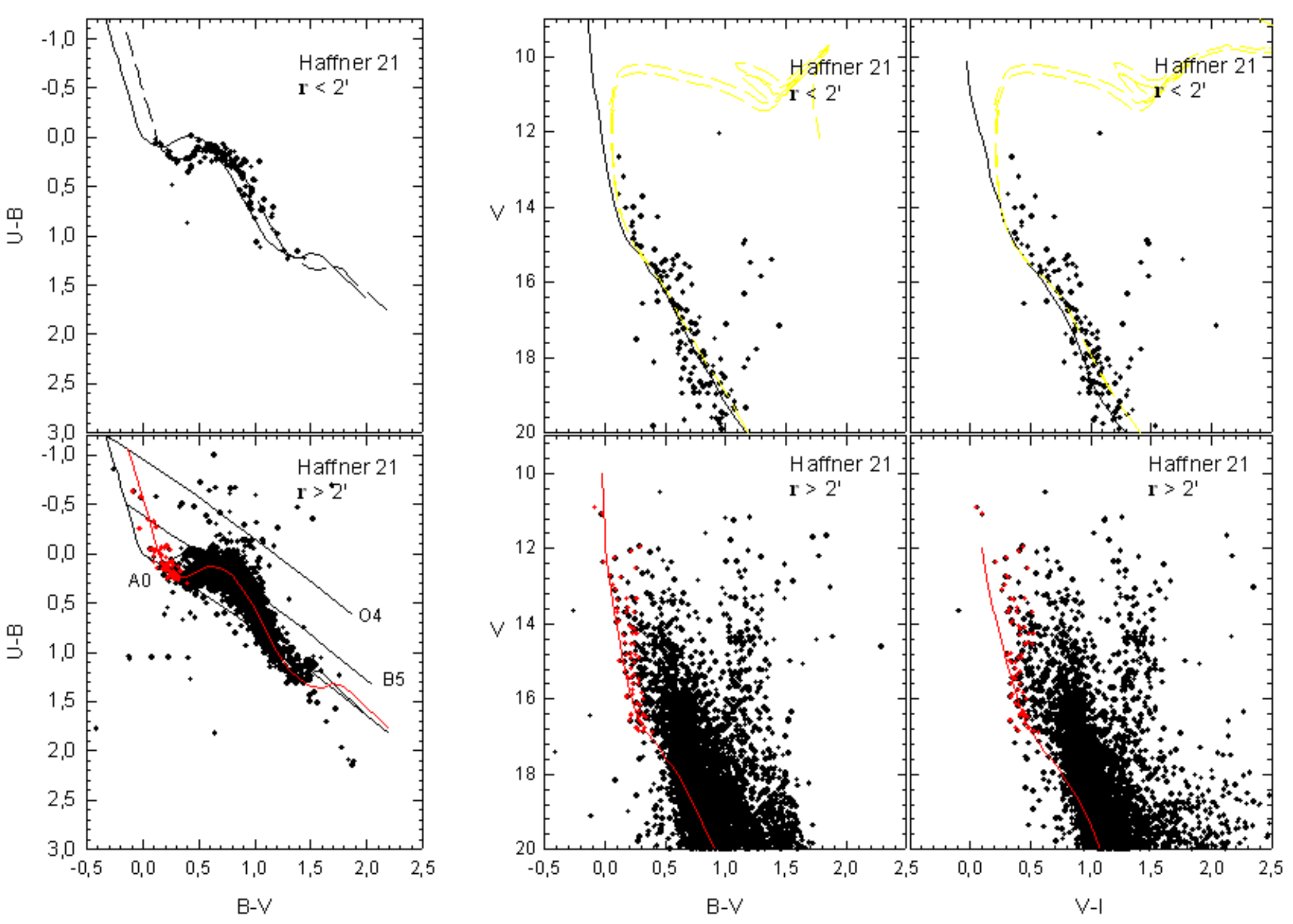} \caption{Color-color and
color-magnitude diagrams for Haffner~21 (upper panels), and for
the surrounding field (lower panels). Dashed lines are isochrone
from Marigo et al. (2008) shifted by color excess and distance
modulus. The solid line in the two color diagram is a
reddening-free ZAMS, while the dashed  line is the same ZAMS
shifted by the color excess to fit the cluster's probable members.
In the lower panels, red colors are used to identify a background
young population, and the ZAMS that better fits its sequence. See
text for details.}
\end{figure}

\clearpage
\begin{figure}
\epsscale{1.} \plotone{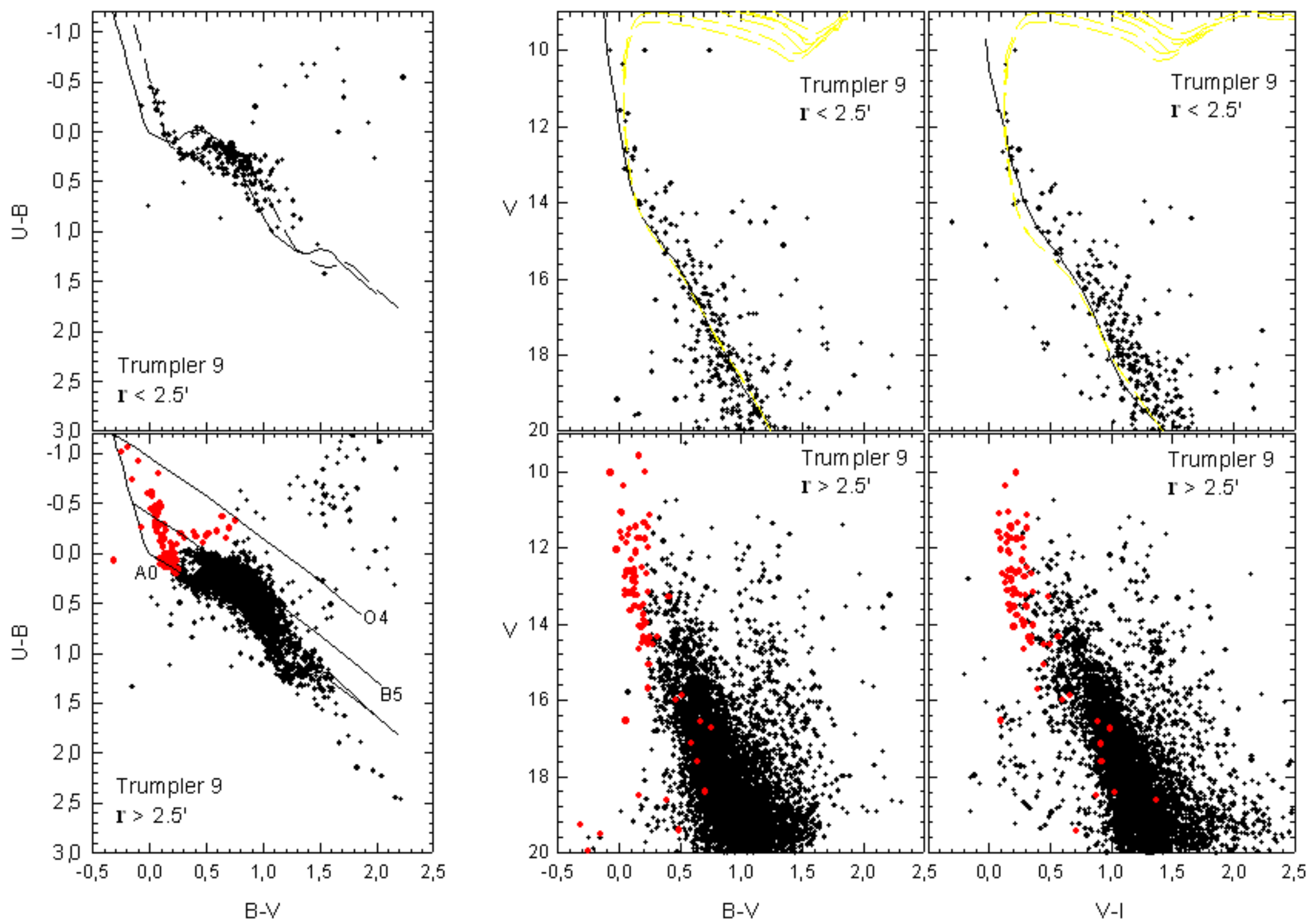} \caption{Color-color and
color-magnitude diagrams for Trumpler~9  (upper panels), and for
the surrounding field (lower panels). Dashed lines are isochrone
from Marigo et al. (2008) shifted by color excess and distance
modulus. The solid line in the two color diagram is a
reddening-free ZAMS, while the dashed  line is the same ZAMS
shifted by the color excess to fit the cluster's probable members.
In the lower panels, red symbols are used to identify a background
young population, and the ZAMS that better fits its sequence. See
text for details. }
\end{figure}

\clearpage
\begin{figure}
\epsscale{1.}
\plotone{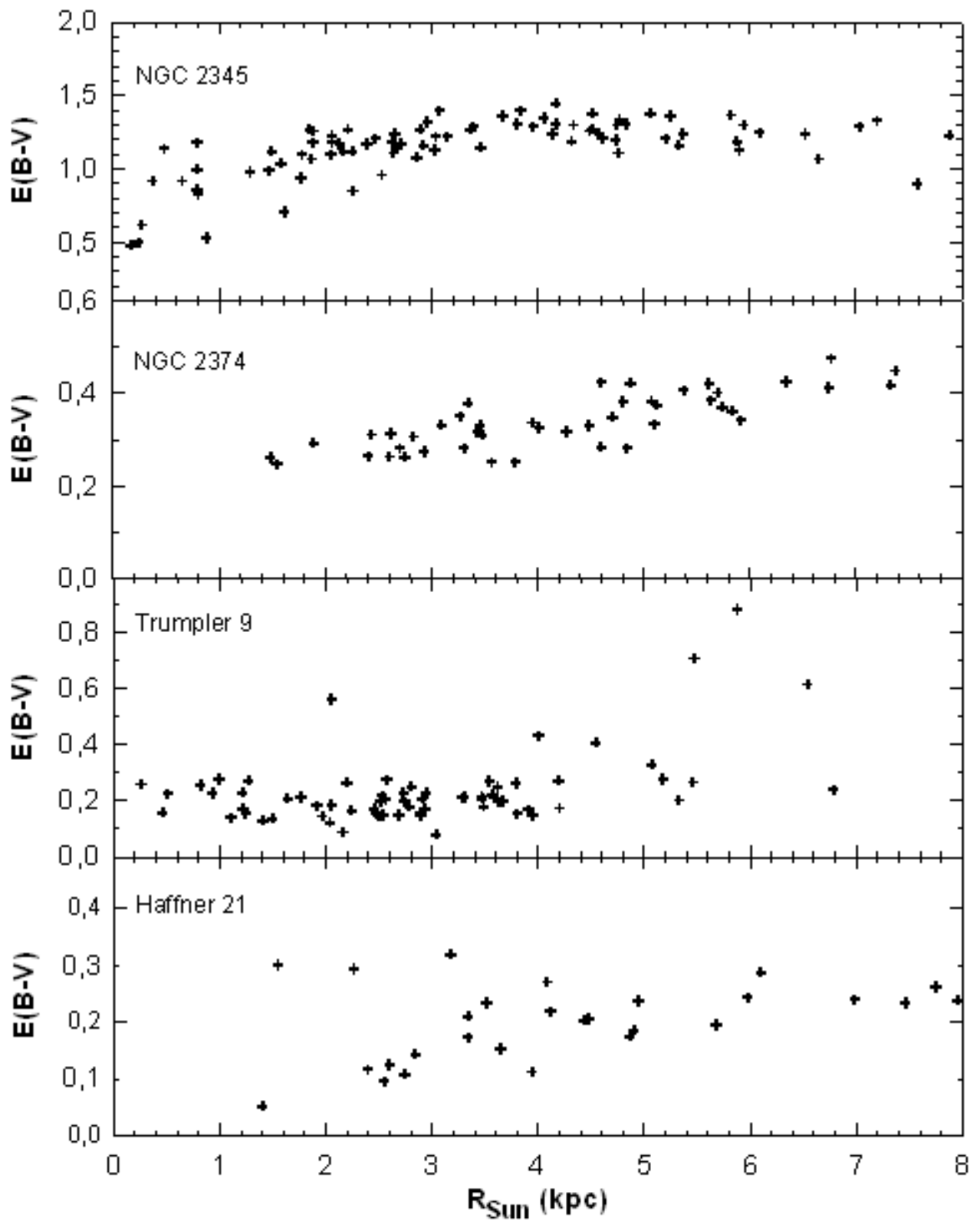}
\caption{Trend of reddening as a function of distance from the Sun for the five directions studied in this work.}
\end{figure}

\clearpage
\begin{figure}
\epsscale{1.} \plotone{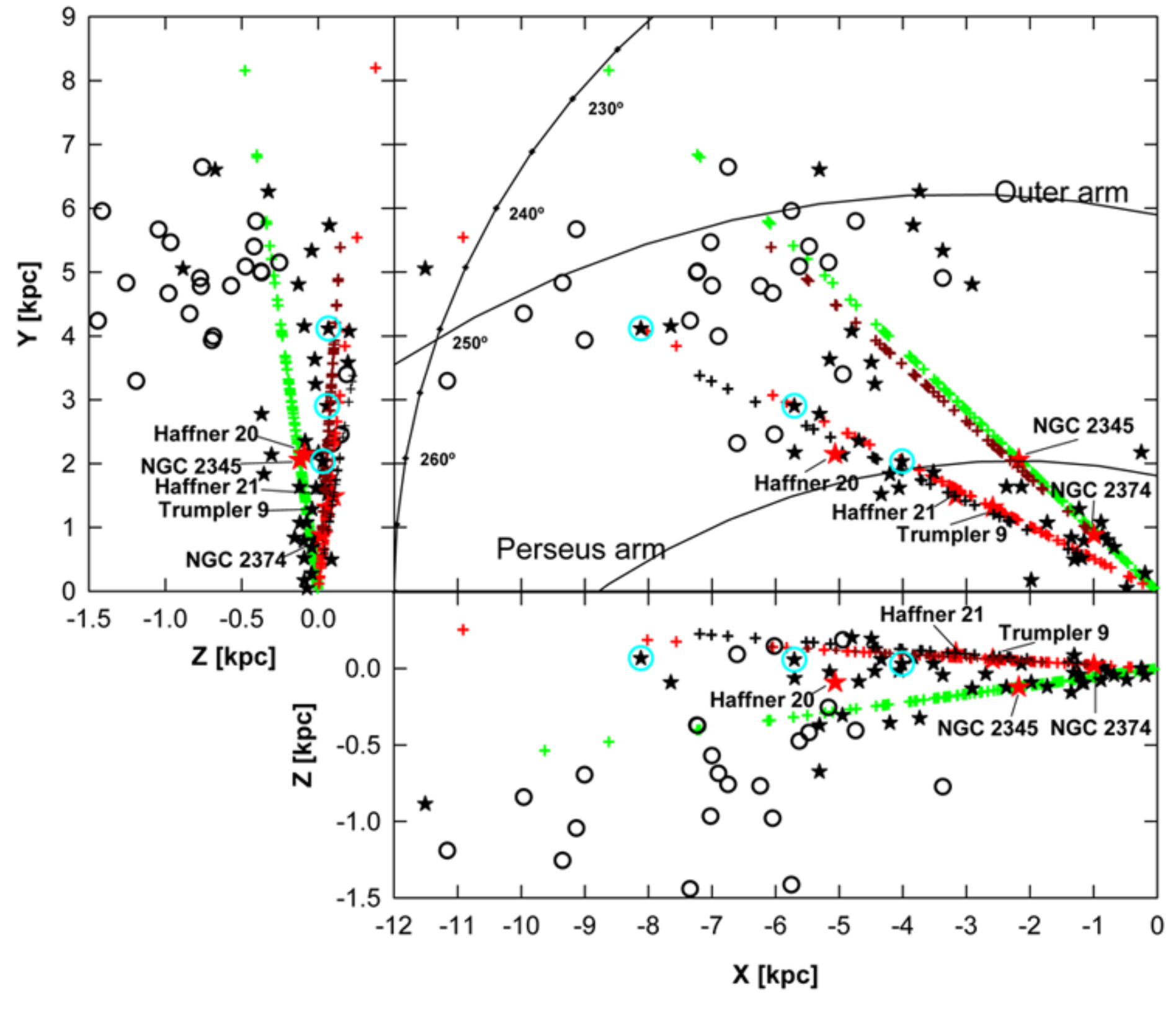} \caption{Updated version of
Fig.~1, where we have added the results of the analysis from the
cluster sample presented in this paper. Symbols are as in Fig.~1,
except that Haffner 18 (1/2)  and 19 are highlighted by cyan circles. Red
stars indicated the five additional clusters studied in this work.
Green, brown, red and black small crosses correspond to
early type stars in the line of sight to NGC~2345, NGC~2374,
Trumpler~9 and Haffner 21, respectively.}
\end{figure}

%%%%%%%%%%%%%%%%%%%%%%%  TABLES %%%%%%%%%%%%%%%%%%%%%%%%%%%%%%%%%%%
\clearpage

\clearpage
%table1
\begin{table}
\caption{Cluster coordinates for J2000.0}
\begin{tabular}{lcccc}
\hline \hline
Cluster & RA & DEC & $l$ & $b$\\
\hline
NGC~2345     & 07:08:18 & -13:11:36 & 226.58 & -2.31 \\
NGC~2374     & 07:23:56 & -13:15:48 & 228.41 &  1.02 \\
Trumpler~9   & 07:55:40 & -25:53:00 & 243.07 &  1.28 \\
Haffner~21   & 08:01:09 & -27:13:00 & 244.85 &  1.63 \\
Haffner~20   & 07:56:15 & -30:22:00 & 246.97 & -0.93 \\
\hline
\end{tabular}
\end{table}

\clearpage
%table3
\begin{deluxetable}{lccccc}
%\tabletypesize{\scriptsize}
\tablecaption{\label{tab:log}%
Log of observations}
\tablewidth{0pt}
\tablehead{
\colhead{Date} &
\colhead{Cluster} & \colhead{Filter} &
\colhead{Exposure Time (sec)} & \colhead{Airmass} & \colhead{Seeing}
}
\startdata
2010 Dec 30 & Haffner 20 & $V$ & 2x10,60,900   & 1.00$-$1.01 & 0.9$-$1.2\\
            &            & $B$ & 20, 3x150     & 1.04$-$1.06  & 0.9$-$1.2\\
            &            & $U$ & 30, 300       & 1.06$-$1.07  & 0.9$-$1.2\\
            &            & $I$ & 3x10,60,900   & 1.02$-$1.03  & 0.9$-$1.2\\
            & Trumpler 9 & $V$ & 2x10,60,900   & 1.00  &0.9$-$1.2\\
            &            & $B$ & 20,150,1500   & 1.12$-$1.19 & 0.9$-$1.2\\
            &            & $U$ & 2x30,300,2000 & 1.00$-$1.03 &0.9$-$1.2\\
            &            & $I$ & 2x10,60,900   & 1.20$-$1.28 & 0.9$-$1.2\\
2010 Dec 31 & Haffner 20 & $U$ & 2x20,2000     & 1.25$-$1.27 & 0.8$-$1.2\\
            &            & $B$ & 20, 1500      & 1.37$-$1.38 & 0.8$-$1.2\\
            & Haffner 21 & $U$ & 30,300,2000   & 1.12$-$1.14 &0.8$-$1.2\\
            &            & $B$ & 20,150,1500   & 1.02$-$1.04 &0.8$-$1.2\\
            &            & $V$ & 10,60,900     & 1.00 &0.8$-$1.2\\
            &            & $I$ & 2x10,60,900   & 1.00$-$1.01 & 0.8$-$1.2\\
2008 Feb 01 & NGC 2345   & $U$ & 5,20,100,200  & 1.09$-$1.10 &1.0$-$1.4\\
            &            & $B$ & 5,20,100,200  & 1.07$-$1.08 &1.0$-$1.4\\
            &            & $V$ & 5,10,60,120   & 1.08$-$1.09 &1.0$-$1.4\\
            &            & $R$ & 5,10,60,120   & 1.09$-$1.12&1.0$-$1.4\\
            &            & $I$ & 5,10,100,200  & 1.06$-$1.07&1.0$-$1.4\\
            & NGC 2374   & $U$ & 5,20,100,200  & 1.11$-$1.12 & 1.0$-$1.4\\
            &            & $B$ & 5,20,100,200  & 1.14$-$1.15 & 1.0$-$1.4\\
            &            & $V$ & 5,10,60,120   & 1.22$-$1.23 &1.0$-$1.4\\
            &            & $R$ & 5,10,60,120   & 1.25$-$1.27 &1.0$-$1.4\\
            &            & $I$ & 5,10,100,200  & 1.17$-$1.19 &1.0$-$1.4\\
2008 Feb 03 & NGC 2345   & $U$ & 60,600,1500   & 1.05$-$1.07 & 0.9$-$1.1\\
            &            & $B$ & 30,600,1500   & 1.05$-$1.08 &0.9$-$1.1\\
            &            & $V$ & 30,600,1200   & 1.08$-$1.09 &0.9$-$1.1\\
            &            & $R$ & 30,600,1200   & 1.09$-$1.12 &0.9$-$1.1\\
            &            & $I$ & 30,600,1200   & 1.07$-$1.08 &0.9$-$1.1\\
            & NGC 2374   & $U$ & 60,600,1500   & 1.06$-$1.07 &0.9$-$1.1\\
            &            & $B$ & 30,600,1500   & 1.11$-$1.13 &0.9$-$1.1\\
            &            & $V$ & 30,600,1200   & 1.31$-$1.38 &0.9$-$1.1\\
            &            & $R$ & 30,600,1200   & 1.50$-$1.60 &0.9$-$1.1\\
            &            & $I$ & 30,600,1200   & 1.19$-$1.24 &0.9$-$1.1\\
\enddata
\end{deluxetable}

\clearpage
\begin{deluxetable}{ccccccc}
\tablecaption{\label{tab:coeficientes}%
Coefficients of the transformation equations}
\tablewidth{0pt}
\tablehead{
\colhead{Mag} &
\colhead{$c_0$} & \colhead{$\sigma$} &
\colhead{$c_1$} & \colhead{$\sigma$} &
\colhead{$c_2$} & \colhead{$\sigma$}
}
\startdata
\cutinhead{%
CTIO, 1-m telescope, 2008 February 01}
$U$& $+1.35$ & 0.01 & $-0.49$ &  0.01 & $+0.015$ & 0.016\\
$B$& $+2.27$ & 0.01 & $-0.25$ &  0.01& $-0.100$ &   0.015\\
$V$& $+2.36$ &  0.01 & $-0.17$ &  0.01 & $-0.010$ &  0.005 \\
$R$& $+2.10$ & 0.01 & $-0.10$ &  0.01& $0.048$ &   0.005\\
$I$& $+1.36$ &  0.01 & $-0.07$ &  0.01 & $-0.043$ &  0.005 \\
\cutinhead{%
CTIO, 1-m telescope, 2008 February 03}
 $U$ & $+1.35$ &0.01 & $-0.49$ &  0.01 & $+0.014$ & 0.020\\
$B$& $+2.20$ & 0.02 & $-0.25$ &  0.01& $-0.097$ &   0.014\\
$V$& $+2.32$ &  0.01 & $-0.16$ &  0.01 & $-0.012$ &  0.006 \\
$R$& $+2.14$ & 0.01 & $-0.09$ &  0.01& $ 0.050$ &   0.010\\
$I$& $+1.35$ &  0.01 & $-0.08$ &  0.01 & $-0.045$ &  0.005 \\
\cutinhead{%
CTIO, 1-m telescope, 2010 December 30}
 $U$ & $+0.89$ &0.02 & $-0.52$ &  0.01 & $+0.019$ & 0.008\\
$B$& $+2.09$ & 0.02 & $-0.27$ &  0.01& $-0.115$ &   0.007\\
$V$& $+2.30$ &  0.01 & $-0.15$ &  0.01 & $+0.016$ &  0.005 \\
$I$& $+1.29$ &  0.01 & $-0.06$ &  0.01 & $-0.045$ &  0.005 \\
\cutinhead{%
CTIO, 1-m telescope, 2010 December 31}
$U$ & $+0.90$ & 0.01 & $-0.50$ &  0.01 & $+0.032$ & 0.006\\
$B$ & $+2.10$ & 0.01 & $-0.25$ &  0.01 & $-0.112$ & 0.005\\
$V$ & $+2.26$ & 0.01 & $-0.11$ &  0.01 & $+0.030$ & 0.005 \\
$I$ & $+1.30$ & 0.01 & $-0.07$ &  0.01 & $-0.048$ & 0.005 \\
\enddata
\end{deluxetable}

\clearpage
\begin{table}
\caption{New and earlier results for the present cluster sample}
\begin{tabular}{lcccccl}
\hline\hline \multicolumn{1}{c}{Cluster} &\multicolumn{1}{c}{Size}
&\multicolumn{1}{c}{$E_{B-V}$} &\multicolumn{1}{c}{$R_{Sun}$}
&\multicolumn{1}{c}{$D$}&\multicolumn{1}{c}{Age}
&\multicolumn{1}{l}{Reference} \\
& \multicolumn{1}{c}{arcmin}& &
\multicolumn{1}{c}{kpc}&\multicolumn{1}{c}{$pc$}&\multicolumn{1}{c}{($ \times 10^6$)~yr}& \\

\hline

NGC 2345    &   5.25    &   0.48--1.16  &   1.75 &        &   60        & Moffat (1974) \\
            &   3.75    &   0.59        &   3.00 &  6.6   &   63--70    &   This work   \\
\hline
NGC 2374    &           &  0.175        &   1.20 &        &   75       &   Babu (1985) \\
            &           &               &        &        &    2000    & Lyng{\aa} (1980)\\
            &           &               &        &        &     350       &    Fenkart et al. (1972)\\
            &           &   0.07        &   1.33 &        &   220-- 280  &   This work\\
\hline
Trumpler 9  &   3.6     &   0.20         &   0.90 &       &               &  Pi{\c{s}}mi{\c{s}}  (1970) \\
            &   3.5     &   0.25        &   2.23  &       &               &   Vogt \& Moffat (1972)  \\
            &   2.5     &   0.20         &   2.90 & 4.2   &   10          &   This work\\
\hline
Haffner 20  &  1.1      &   0.55        &   2.40  &       &   200         &   FitzGerald \& Moffat (1974)  \\
            &   $<$ 2.0 &   0.65        &   5.50  & 6.2   &   90-- 100    &   This work   \\
\hline
Haffner 21  &   1.1     &   0.20        &   3.30  &       &   200    &   FitzGerald \& Moffat (1974)  \\
            &   $<$2.0  &   0.21        &   3.50  & 4.0   &   110-120            &   This work   \\
\hline
\end{tabular}
\end{table}


\begin{thebibliography}{}
\bibitem[Aarseth (1996)]{aar96} Aarseth, S. J., 1996, in The Origins, Evolution, and Destinies of Binary Stars in Clusters, ed. E. F. Milone, \& J.-C. Mermilliod, ASP Conf. Ser., 90, 423
\bibitem[Alfaro et al. (1991)]{alf91} Alfaro, E.J., Cabrera-Cano, J., Delgado, A.J., 1991, ApJ, 378, 106
\bibitem[Anderson et al. (2014)]{and14} Anderson, L. D.,  Bania, T. M.,  Balser, D. S., Cunningham, V., Wenger, T. V., Johnstone, B. M., Armentrout, W. P., 2014, ApJS, 212, 1
\bibitem[Babu (1985)]{bab85} Babu, S., 1985,  Journal of Astrophys. Astr., 6, 61
\bibitem[Baume et al. (2004)]{baum04} Baume, G., Moitinho, A., Giorgi, E. E., Carraro, G., \& V\'azquez, R. A. 2004, A\&A, 417, 961
\bibitem[Breedt \& Gansicke (2011)]{bee11} Breedt, E., Gansicke, B.T., 2011, ASPC, 447, 203
\bibitem[Carraro et al. (2005)]{ca05} Carraro, G., V\'azquez, R.A., Moitinho, A., Baume, G., 2005, ApJ, 630, L153
\bibitem[Carraro et al. (2007)]{car07} Carraro, G., Moitinho, A., Zoccali, M., V\'azquez, R.A.,  Baume, G., 2007, AJ, 133, 1058
\bibitem[Carraro et al. (2010a)]{ca10a} Carraro, G., Costa, E., \& Ahumada, J. A. 2010a,    AJ, 140, 954
\bibitem[Carraro (2011)]{car11} Carraro, G., 2011, A\&A, 536, 101
\bibitem[Carraro et al. (2010b)]{ca10b} Carraro, G., V\'azquez, R.A., Costa, E., Perren, G., Moitinho, A., 2010b, ApJ, 718, 683
\bibitem[Carraro et al. (2013)]{car13} Carraro, G., Turner, D., Majaess, D., Baume, G., 2013, A\&A, 555, 50
\bibitem[Carraro et al. (2014)]{carr14} Carraro, G., Perren, G., V\'azquez, R.A., Moitinho, A.,
in "Structure and Dynamics of Disk Galaxies" , 2014, M. Seigar and P. Treuthardt, eds, ASPC, 480, 10
\bibitem[Carraro (2014)]{car14} Carraro, G., Proceedings of the IAU Symposium No. 298, "Setting the scene for Gaia and LAMOST", eds. S. Feltzing, G. Zhao, N. A. Walton, and P. A. Whitelock, 2014, IAUS 298, 7
\bibitem[de La Fuente (1997)]{delaf97} de La Fuente Marcos, R. 1997, A\&A, 322, 764
\bibitem[Downes (1986)]{do86} Downes, R.A., 1986, ApJS, 61, 569
\bibitem[Fenkart et al. (1972)]{fen72} Fenkart, R.P., Buser, R., Ritter, H., Schmitt, H., Steppe, H., Wagner, R., Weidemann, D.,  1972, A\&AS, 7, 48
\bibitem[Fitzgerald \& Moffat (1974)]{fm74} Fitzgerald, M. P., \& Moffat, A. F. J. 1974, PASP, 86, 480
\bibitem[Fujii \& Portegies Zwart (2011)]{fu11} Fujii, M.S., Portegies Zwart, S., 2011, Science, 334, 1380
\bibitem[Havlen (1976)]{hav76} Havlen R. J., 1976, A\&A, 47, 193
\bibitem[Humphreys (1978)]{hum78} Humphreys R. M., 1978, ApJS, 38, 309
\bibitem[Kaltcheva \& Hilditch (2000)]{kalt00} Kaltcheva, N.T., \& Hilditch, R.W., 2000, MNRAS 312, 753
\bibitem[King (1962)]{king62}King, I. 1962, AJ, 67, 471
\bibitem[Kroupa et al. (2001)] {kroup01} Kroupa, P., Aarseth, S., \& Hurley, J. 2001, MNRAS, 321, 699
\bibitem[Landolt (1992)]{lan92} Landolt, A. U. 1992, AJ, 104, 372
\bibitem[Levine et al. (2006)]{lev06} Levine, E.S., Blitz, L., Heiles, C., 2006, Science, 312, 1773
\bibitem[Lyng{\aa} (1980)]{lyn80} Lyng{\aa}, G. 1980, A Computer Readable Catalogue of Open Cluster
    Data, Stellar Data Centre, Observatoire de Strasbourg, France
 \bibitem[L\'opez-Corredoira \& Molg\'o (2014)]{lo14} L\'opez-Corredoira, M., Molg\'o, J, 2014, A\&A, 567, 106
\bibitem[Marigo et al. (2008)]{mar08} Marigo, P., Girardi, L., Bressan, A., Groenewegen, M.A.T., Silva, L., Granato, G.L., 2008, A\&A, 482, 883
\bibitem[Martin et al. (2004)]{mart04} Martin N., Ibata R.A., Bellazzini M., Irwin M.J., Lewis G.F., Dehnen W., 2004, MNRAS 348, 12
\bibitem[Moffat (1974)]{mof74} Moffat, A.F.M.,  1974, A\&AS 16, 33
\bibitem[Moitinho (2001)]{moit2001} Moitinho, A., 2001, A\&A 370, 436
\bibitem[Moitinho et al. (2006)]{Moi06} Moitinho, A., V\'azquez, R.A., Carraro, G., Baume, G., Giorrgi, E.E., Lyra, W., 2066, MNRAS, 368, L77
\bibitem[Momany et al. (2006)]{mom06} Momany, Y., Zaggia, S., Gilmore, G., Piotto, G., Carraro, G., Bedin, L.R., de Angeli, F., 2006, A\&A, 451, 515
\bibitem[Perren et al. (2012)]{per12} Perren, G., V\'azquez, R.A., Carraro, G., 2012, A\&A, 548, 125
\bibitem[Pi{\c{s}}mi{\c{s}}(1970)]{pis70} Pi{\c{s}}mi{\c{s}}, P. 1970, Bol. Obs. Tonantzintla y Tacubaya, 5, 293
\bibitem[Robin et al. (1992)]{rob92} Robin, A., Creze, M., Mohan, V., 1992, ApJ, 400, L25
\bibitem[Roman-Lopes et al. (2011)]{rom11} Roman-Lopes, A., Barba, R.H., Morrell, N.I., 2011, MNRAS, 416, 501
\bibitem[Schmidt-Kaler (1982)]{sch82}  Schmidt-Kaler, Th., 1982 , Landolt-B\"ornstein, Numerical
    data and Functional Relation\-ships in Science and Technology,
    New Series, Group VI, Vol. 2(b), K. Schaifers \& H. H. Voigt, eds.,
    Springer-Verlag, Berl\'{\i}n, p. 14
\bibitem[Skrutskie et al.(2006)]{skr06} Skrutskie, M. F., Cutri, R. M., Stiening, R., et al. 2006, AJ, 131, 1163
\bibitem[Stetson (1987)]{ste87} Stetson, P. B. 1987, PASP, 99, 191
\bibitem[Strai\v{z}ys (1991)]{str91} Strai\v{z}ys, V., 1991, in Multicolor Photometry, Astronomy and Astrophysics series 15, Pachart Publishing House
\bibitem[Suad et al. (2014)]{su14} Suad, L.A., Caiafa, C.F., Arnal, W.M., Cichowolski, S., 2014, A\&A, in press
\bibitem[Vall{\'{e}}e (2008)]{val08} Vall\'ee, J. P. 2008, AJ, 135, 1301
\bibitem[V\'azquez et al. (2008)]{vaz08} V\'azquez, R.A., May, J., Carraro, G., Bronfman, L., Moitinho, A., Baume, G., 2008, ApJ, 672, 930
\bibitem[V\'azquez et al. (2010)]{vaz10} V\'azquez, R.A., Moitinho, A., Carraro, G., Dias, W.S., 2010, A\&A, 511, 38
\bibitem[Vogt \& Moffat (1972)]{vo72} Vogt N., Moffat, A.F.J., 1972, A\&AS, 7, 133

\end{thebibliography}
\end{document}